\documentclass[usenatbib,useams]{mn2e}
\usepackage{times, aas_macros, graphicx}
\usepackage{mathrsfs}

\pdfoutput=1
\defcitealias{vdm2012}{vdM12}
\defcitealias{courteau99}{CB99}

\def \Msun{ {\rm M}_{\sun} }
\def \kms{ {\rm km ~ s^{-1}} }

\def \mrat{ M_{\rm M31}/M_{\rm MW} }

\def \vlos { V_{\rm los} }

\def \vn { V_{\rm N} }
\def \vw { V_{\rm W} }

\def \vni{ V_{{\rm N} i} }
\def \vwi{ V_{{\rm W} i} }

\def \lmw{ l_{\textsc{\tiny \rm MW}} }
\def \bmw{ b_{\textsc{\tiny \rm MW}} }

\def \vsun{ V_{\odot} }
\def \lsun{ l_{\odot} }
\def \bsun{ b_{\odot} }

\def \vmw{ V_{\rm MW} }
\def \vand{ V_{\rm M31} }
\def \vcir{ V_{\rm cir} }

\def \los{ \bmath{\hat{\ell}} }
\def \east{ \bmath{\hat{e}} }
\def \north{ \bmath{\hat{n}} }

\def \rhat{ \bmath{\hat{r}} }
\def \that{ \bmath{\hat{\theta}} }
\def \phat{ \bmath{\hat{\phi}} }

\def \u{ \bmath{\hat{u}} }

\def \angmm{ \delta_{\textsc{\tiny \rm MW,M31}} }

\def \vsunlg{ \bmath{V}_{\sun \rightarrow {\rm LG} } }
\def \vmwlg{ \bmath{V}_{{\rm MW} \rightarrow {\rm LG} } }
\def \vandlg{ \bmath{V}_{{\rm M31} \rightarrow {\rm LG} } }
\def \vmwsun{ \bmath{V}_{{\rm MW} \rightarrow \sun } }
\def \vandsun{ \bmath{V}_{{\rm M31} \rightarrow \sun } }
\def \vandmw{ \bmath{V}_{{\rm M31} \rightarrow {\rm MW} } }

\title[Balancing mass and momentum in the Local Group]{Balancing mass and momentum in the Local Group}

\author[J. D. Diaz et al.]{
   J. D. Diaz,$^1$\thanks{Email: jdiaz@ast.cam.ac.uk}
   S. E. Koposov,$^1$ M. Irwin,$^1$ V. Belokurov,$^1$ and N. W. Evans$^1$ \\
   $^1$Institute of Astronomy, Madingley Road, Cambridge CB3 0HA 
}

\date{}

\begin{document}

\maketitle
\begin{abstract}
In the rest frame of the Local Group (LG), the total momentum of the Milky Way (MW) and Andromeda (M31) should balance to zero.  We use this fact to constrain new solutions for the solar motion with respect to the LG centre-of-mass, the total mass of the LG, and the individual masses of M31 and the MW.  Using the set of remote LG galaxies at $>350$ kpc from the MW and M31, we find that the solar motion has amplitude $V_{\odot} = 299 \pm 15 {\rm ~km ~ s^{-1}}$ in a direction pointing toward galactic longitude $l_{\odot}= 98.4^{\circ} \pm 3.6^{\circ}$ and galactic latitude $b_{\odot}=-5.9^{\circ}\pm 3.0^{\circ}$.  The velocities of M31 and the MW in this rest frame give a direct measurement of their mass ratio, for which we find $\log_{10} (M_{\rm M31}/M_{\rm MW})=0.36 \pm 0.29$.  We combine these measurements with the virial theorem to estimate the total mass within the LG as $M_{\rm LG} = (2.5 \pm 0.4) \times 10^{12}~\Msun$.

Our value for $M_{\rm LG}$ is consistent with the sum of literature values for $M_{\rm MW}$ and $M_{\rm M31}$.  This suggests that the mass of the LG is almost entirely located within the two largest galaxies rather than being dispersed on larger scales or in a background medium. \textit{The outskirts of the LG are seemingly rather empty.}  Combining our measurement for $M_{\rm LG}$ and the mass ratio, we estimate the individual masses of the MW and M31 to be $M_{\rm MW}=(0.8 \pm 0.5) \times 10^{12}~\Msun$ and $M_{\rm M31}=(1.7 \pm 0.3) \times 10^{12}~\Msun$, respectively.  Our analysis favours M31 being more massive than the MW by a factor of $\sim$2.3, and the uncertainties allow only a small probability (9.8\%) that the MW is more massive.  This is consistent with other properties such as the maximum rotational velocities, total stellar content, and numbers of globular clusters and dwarf satellites, which all suggest that $M_{\rm M31}/M_{\rm MW}>1$.

\end{abstract}


\section{Introduction} \label{sec:int}

The nearby Andromeda galaxy (M31) and our own Milky Way (MW) are the main members of the Local Group (LG) of galaxies, with the other members being separated into one of two categories: the cadre of dwarf satellites accompanying each spiral, and the dozen or so independent galaxies sprinkled in space up to $\sim$1.5 Mpc away.  The boundary of the LG is not precise, but rather is marked by those galaxies which are unambiguously participating in cosmic expansion (e.g. \citealt{kara2009,vdb99ant}).  One may in fact define the LG as the self-bound set of galaxies which have detached from the Hubble flow.

Though it is well known that the luminosity of the LG is dominated by the MW and M31 (\citealt{vdb99}), the distribution of mass within the LG is poorly understood in comparison.  Methods to estimate the dynamical mass of the LG traditionally assume that the halos of M31 and the MW are the only reservoirs of mass (e.g. \citealt{dlb81, edlb82, sandage86, kara2005, vdm2008}).  That assumption comes in stark contrast with $\Lambda$CDM simulations, where the remnants of accretion generate a massive background medium in the group (\citealt{cox2008, gonzalez2013}).  Despite the wealth of data on the galaxies of the LG (see \citealt{mcconn2012} and references therein), the constraints on the existence of such a background medium are nonexistent.  In fact,  the state of affairs regarding our knowledge of mass within the LG is so limited that there is still no consensus on whether the MW or M31 is more massive.

The range of recent mass estimates for M31 has a lower bound of $\sim$0.8$ \times 10^{12}~ \Msun$ (\citealt{tamm2012, seigar2008, brunthaler2007}), an upper bound of roughly $2 \times 10^{12}~ \Msun$ (\citealt{fardal2013, leemg2008, galleti2006}), and a smattering of intermediate estimates (\citealt{veljanoski2013, watkins2010, corbelli2010, tollerud2012, evans2003, klypin2002}).  A similarly uncertain range of $0.5-2 \times 10^{12}~ \Msun$ applies for recent estimates of the total MW mass (\citealt{gibbons2014, piffl2014, mbk2013, alis2012mass, busha2011, mcmillan2011, watkins2010, xue2008, klypin2002}).  In addition to the large scatter among these mass inferences, all of the aforementioned studies differ in their assumptions, systematics, adopted kinematic tracers, and range of valid radii.  Compounding the uncertainty even further, these figures would appear to be underestimates in light of the timing argument, which places the sum of the M31 and MW masses at $5 \times 10^{12}~ \Msun$ (\citealt{partridge2013, vdm2012, li&white2008}).

Because they are the most distant kinematic tracers, the dwarf satellites offer the most promising opportunity to measure the total masses of M31 and the MW in a consistent manner.  However, the latest statistical analysis of the satellite data is unable to discern whether the MW or M31 is more massive, with each spiral assigned a mass of roughly $1.4 \pm 0.4 \times 10^{12} ~\Msun$ within 300 kpc (\citealt{watkins2010}).  Additional complications, such as the uncertainty arising from velocity anisotropies and the choice to include/exclude various satellites, cause the mass ratio to be largely unconstrained.

In this paper our first goal is to directly measure the relative mass of M31 and the MW.  Our method can be summarized in two steps: first, we use the kinematics of the outlying members of the LG to determine the rest frame of the LG; and second, we measure the velocities of M31 and the MW within this rest frame.  The ratio of these velocities gives us an estimate of the relative mass.  The second goal of the present work is to estimate the total mass of the LG using the same set of outer LG galaxies.  We first compute the kinetic and potential energies of these remote galaxies using our solution for the LG rest frame, and then we employ the virial theorem to extract a mass estimate.

In the literature there is a well-established method for measuring the rest frame of the LG from the collective motions of its members (\citealt{yahil77, kara96, vdb99, courteau99}).  In the present work we improve upon the results of previous authors in three important ways.  First, we remove the influence of satellite motions from the solution, because the satellites diminish the influence of the outer LG members while over-representing the MW and M31.  Second, whereas previous treatments have utilized line-of-sight data alone, we are able to make use of measured transverse velocities.  And third, we ensure that our solution for the LG rest frame is consistent with the balance of linear momentum between the MW and M31.

The balance of momentum within the LG has been considered previously (e.g. \citealt{edlb82}), but our study is the first to utilize this constraint to measure the relative mass of M31 and the MW.  Our analysis relies on the recently measured transverse velocity of M31 (\citealt{sohn2012}) and the latest insights into the circular velocity of the MW (\citealt{reid2009, mcmillan2011}).

The structure of our paper is as follows.  In Section \ref{sec:tbg}, we introduce the equations which underpin our study, including the balance of linear momentum.  We also present the vector notation that we adopt throughout this work, and we provide our choice of parameters.  In Section \ref{sec:meh}, we outline our statistical procedure for measuring the solar motion with respect to the LG centre-of-mass (Step 1, Section \ref{sec:meh1}), and for measuring the relative mass of M31 and the MW (Step 2, Section \ref{sec:meh2}).  The results of that analysis are given in Section \ref{sec:res}, and our solution for the LG rest frame is explored in more detail in Section \ref{sec:lgrf}.  In Section \ref{sec:lgmass} we use the virial theorem to estimate the mass of the LG, and we combine this with our result for the mass ratio to estimate the masses of M31 and the MW.  In Section \ref{sec:dis} we consider the impact of different parameter values on our analysis, and we present a number of supplemental arguments to help us decide whether the MW or M31 is more massive.  Lastly we summarize in Section \ref{sec:sum}.

\section{Theory and Background} \label{sec:tbg}

Throughout this paper we will make use of velocity vector notation with the following arrow subscripts (\citealt{courteau99}):  given objects A and B, the velocity of A with respect to B is written as $\bmath{V}_{{\rm A} \rightarrow {\rm B} }$.  This choice of notation clarifies and simplifies vector manipulations, as seen in the following two identities: $\bmath{V}_{{\rm A} \rightarrow {\rm B} } = -\bmath{V}_{{\rm B} \rightarrow {\rm A} }$, and given another object C, 
$\bmath{V}_{{\rm A} \rightarrow {\rm C} } = \bmath{V}_{{\rm A} \rightarrow {\rm B} } + \bmath{V}_{{\rm B} \rightarrow {\rm C} }$.

\subsection{The Balance of Momentum} \label{sec:tbg1}

Given that the LG is sufficiently isolated from comparably massive nearby groups (\citealt{kara2013}), the internal dynamics of its members determines a natural frame of rest.  By construction, the total linear momentum balances to zero in this rest frame:
\begin{equation} \label{eq:mom0}
   0 = \int \rho ~ \bmath{v} ~ {\rm d}^3 r,
\end{equation}
where the velocities $\bmath{v}$ refer to the LG rest frame and $\rho$ is the mass density within the LG.  As explained below, modest assumptions on $\rho$ and $\bmath{v}$ allow this relation to take the tractable form of
\begin{equation} \label{eq:mom}
   0 = M_{\rm MW} \vmwlg + M_{\rm M31} \vandlg
\end{equation}
where $M_{\rm MW}$ and $M_{\rm M31}$ are the masses of the halos, and $\vmwlg$ and $\vandlg$ are their velocities in the LG.

The easiest way to derive (\ref{eq:mom}) is to take the MW and M31 halos as the only significant contributors to $\rho$.  The halos can be extended (but not overlapping) and the orbit need not take an analytic form.  Compare this to the timing argument (\citealt{dlb81, kochanek96}) which assumes point masses and Keplerian orbits.  Satellites such as the LMC and M33 are ignored to first order because their masses are at least an order of magnitude smaller than the MW and M31 (e.g. \citealt{guo2010}).  Even so, we may simply include the mass of all satellites into the total halo masses $M_{\rm MW}$ and $M_{\rm M31}$.

Equation (\ref{eq:mom}) also applies more generally, because it may hold even if the LG contains a massive intragroup medium (\citealt{cox2008}).  We only require the medium to be static ($\bmath{v}=0$) in the center-of-mass frame, such that it contributes negligible overall momentum to the balance of (\ref{eq:mom0}).  Certainly this static condition will not be true near the halos, due to two important interactions: the halos will sweep up infalling material from the medium, and the halos will transfer orbital energy and angular momentum to the medium via dynamical friction (\citealt{cox2008}).  Nevertheless, we can consider such activity at the fringes of a halo to be a part of the halo itself, which may allow equation (\ref{eq:mom}) to hold at least approximately at each point in the orbit.  This situation would test the applicability of the timing argument, however, because the orbit becomes non-Keplerian as it decays toward a merger.

\subsection{Velocity Decomposition} \label{sec:tbg2}

In this study we will measure $\vmwlg$ and $\vandlg$ and thereby estimate $\mrat$ via equation (\ref{eq:mom}).  To make progress toward this goal, the velocities in the LG rest frame are decomposed as
\begin{eqnarray} \label{eq:decomp}
   \vmwlg &=& \vmwsun + \vsunlg \nonumber \\
   \vandlg &=& \vandsun + \vsunlg,
\end{eqnarray}
where $\vmwsun$ and $\vandsun$ are the heliocentric velocities of the MW and M31, respectively, and $\vsunlg$ is the solar motion with respect to the LG centre-of-mass.  As explained in Section \ref{sec:meh1}, a statistical procedure can be used to estimate $\vsunlg$ from the collective motions of LG member galaxies.  The remainder of this section is devoted to the heliocentric velocities of the spirals.

The heliocentric velocity of the Milky Way is trivially related to the motion of the Sun in the Galactic Standard of Rest (GSR) frame.  We simply write
\begin{equation} \label{eq:vmw.wrt.sun}
   \vmwsun = -\bmath{V}_{\sun \rightarrow {\rm MW} } = -(U_0,~ V_0 + \vcir,~ W_0),
\end{equation}
where $U_0$, $V_0$, and $W_0$ are the components of the Sun's velocity with respect to the Local Standard of Rest (LSR), and $\vcir$ is the circular rotation velocity of the MW at the solar radius.  We adopt values of $(U_0, ~V_0, ~W_0) = (11.1,~ 12.24,~ 7.25) ~\kms$ as measured by \citet{schonrich2010}, with the uncertainties $(1.23,~ 2.05,~ 0.62) ~\kms$ from the quadrature sum of statistical and systematic errors.  While there is much debate on the value of $\vcir$, we will adopt the value $239 \pm 5 ~\kms$ advocated by \citet{mcmillan2011}, which combines the latest measurements to constrain this quantity (\citealt{reid2009, reid&brunthaler, gillessen2009}).  In Section \ref{sec:relax} we consider how a different value for $\vcir$ affects our analysis, and we find it has negligible impact.

By using equation (\ref{eq:vmw.wrt.sun}), we implicitly adopt a coordinate system with the following $x$, $y$, and $z$ directions, centered on the Sun: $x$ is the direction toward the Galactic center (galactic longitude $l=0^{\circ}$ and galactic latitude $b=0^{\circ}$); $y$ is the direction of Galactic rotation $(l=90^{\circ},~b=0^{\circ})$; and $z$ is the direction vertically upward from the Galactic disk $(b=90^{\circ})$.

Given this coordinate system, the unit vector $\los$ along the line-of-sight toward an object (as seen from the Sun) has the parametrization
\begin{equation} \label{eq:los}
   \los = ( \cos b \cos l ,~ \cos b  \sin l,~  \sin b),
\end{equation}
where $l$ is the galactic longitude of the object and $b$ is the galactic latitude.

The heliocentric velocity of M31 can be expressed as the following linear combination of its line-of-sight velocity $\vlos$ and its transverse velocities north $\vn$ and west $\vw$ in the plane of the sky:
\begin{equation} \label{eq:vm31.wrt.sun}
   \vandsun = \vlos \los - \vw \east + \vn \north,
\end{equation}
where $\los$ points from the Sun toward M31 along the line of sight, $\east$ points in the direction eastward (i.e. increasing right ascension) at the sky location of M31, and likewise, $\north$ points in the direction northward (i.e. increasing declination).

The value of $\vlos$ for M31 has long been known from spectroscopic measurements, and we adopt the value $-300 \pm 4 ~\kms$ (\citealt{mcconn2012}).  In contrast, the M31 transverse velocity has only recently been measured, using high-precision astrometry with the \textit{Hubble Space Telescope} (\citealt{sohn2012}).  Combining this astrometric value with other indirect inferences, \citet{vdm2012} derive an unprecedented constraint on the M31 transverse motion.  However, we cannot simply take their advocated values for $\vn$ and $\vw$ because one of their inferences relies on an assumed value for $\mrat$.  Because we intend to measure $\mrat$ in the present work, we must exclude the corresponding inference and re-calculate the weighted averages\footnote{Table 3 of \citet{vdm2012} lists the values for $\vn$ and $\vw$ from three astrometric fields and four indirect inferences.  We exclude the entry ``Outer LG galaxies" since it relies on an assumption for $\mrat$, and we take the weighted average of the remaining data.}.  We get $\vn = -61.9 \pm 33.8 ~\kms$, and $\vw = -119.2 \pm 36.3 ~\kms$.

Even though our adopted values for $\vn$ and $\vw$ differ slightly from \citet{vdm2012}, the main conclusion of that paper remains: to within 1$\sigma$, the data is consistent with a radial MW-M31 orbit.  To illustrate this point, let us construct the unit vectors $\rhat$, $\that$, and $\phat$, where $\rhat$ gives the radial direction (i.e. pointing from the centre of the MW to M31), and the tangential directions are given by $\that$ and $\phat$, where $\phat$ lies in the Galactic plane.  As a set, $\{\rhat,~ \that,~ \phat \}$ form a mutually orthogonal right-handed basis and have the coordinates
\begin{eqnarray} \label{eq:rpt}
   \rhat &=& (-0.4896,~  0.7915,~ -0.3657) \nonumber \\
   \that &=& (0.1924,~ -0.3110,~ -0.9307) \nonumber \\
   \phat &=& (-0.8505,~ -0.5260,~  0).
\end{eqnarray}
This is a particularly useful basis for studying the relative motion of M31 and the MW, and we will make use of it throughout the paper when representing vector quantities.  According to our choice of parameters, we calculate the projection of the relative velocity vector $ \vandmw = \vandsun - \vmwsun$ along these unit vectors as
\begin{eqnarray}
   \rhat \cdot \vandmw &=& -108.2 \pm 5.9 ~\kms \nonumber \\
   \that \cdot \vandmw &=& -17.7 \pm 33.8 ~\kms \nonumber \\
   \phat \cdot \vandmw &=& -21.6 \pm 36.5 ~\kms,
\end{eqnarray}
where each component is Gaussian distributed and is summarized by the mean and standard deviation.

From the above, it is clear that the tangential velocity of M31 with respect to the MW (i.e. along $\that$ and $\phat$) is consistent with zero.  The 1$\sigma$ errors are quite large, however, so many more orbits than simply the radial case are possible.

\section{Method} \label{sec:meh}

In this section we outline our statistical methods for measuring $\vsunlg$ in Step 1 and $\mrat$ in Step 2.

\subsection{Step 1: Measuring $\bmath{\vsunlg}$} \label{sec:meh1}

\begin{table}
   \begin{center}
   \caption{Chosen sets of galaxies within the Local Group.  Selection is based on cuts on distance $D$ from M31 and the MW.  That is, $D<1.5$ Mpc is shorthand for the two conditions $D_{\rm MW}<1.5$ Mpc and $D_{\rm M31}<1.5$ Mpc as applied to the \citet{mcconn2012} catalog.}
   \begin{tabular}{l c c c}
   \hline
   Set & Description & Selection of members & $N$ \\
   \hline
   A            & All galaxies  & $D<1.5$ Mpc & 74 \\
   B$^\dagger$  & No satellites & M31, MW, and $350~{\rm kpc}<D<1.5$ Mpc & 17 \\
   \hline
   \multicolumn{4}{l}{$^\dagger$Denotes the preferred set in our analysis.}
   \label{tab:obj}
   \end{tabular}
   \end{center}
\end{table}

\begin{figure}
   \begin{center}
   \includegraphics[width=0.45\textwidth]{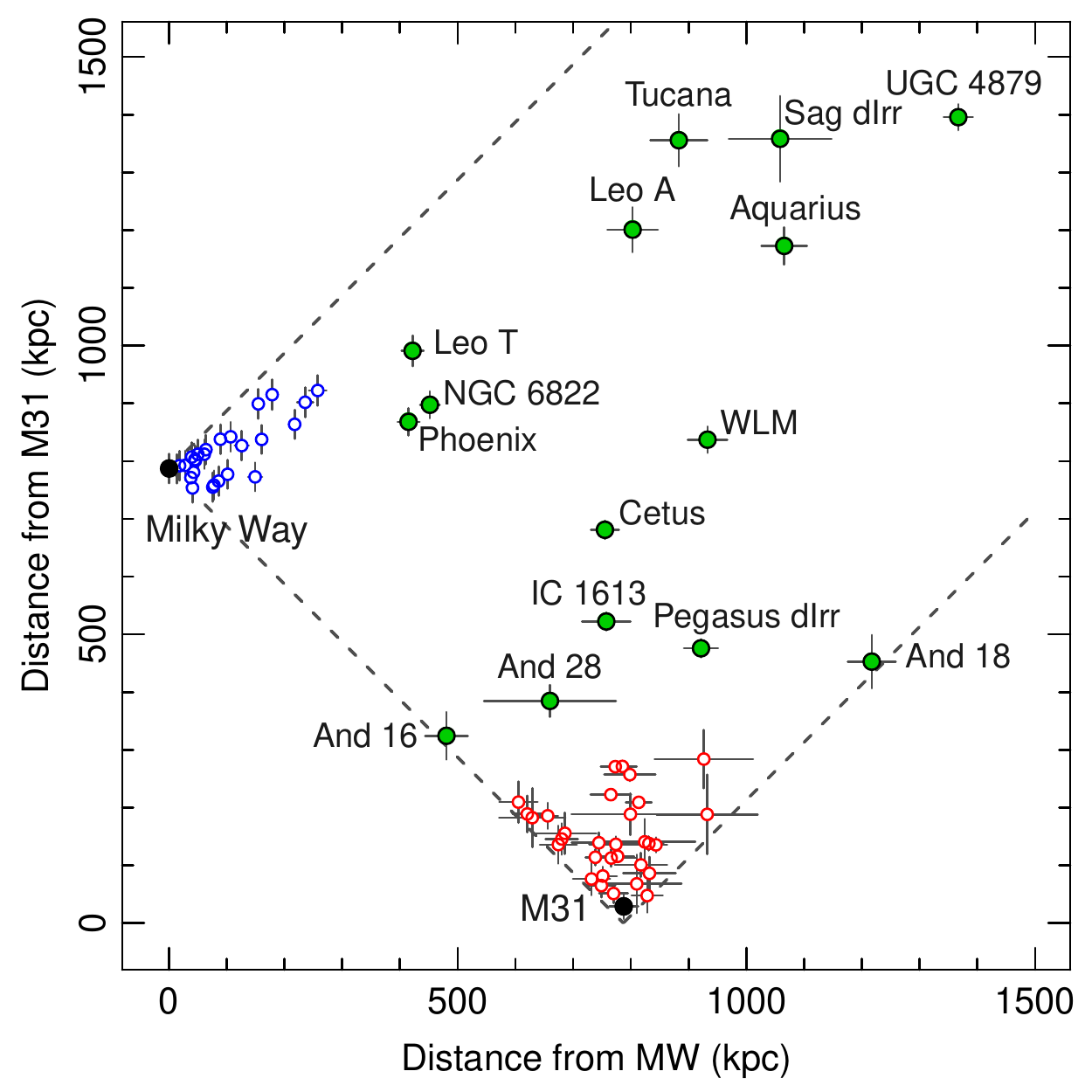}
   \caption{The spatial distribution of LG galaxies taken from the \citet{mcconn2012} catalog, depicted as distance from M31 against distance from the MW.  The grey dashed lines are the boundaries within which the galaxies may be located, and the error bars give the 1$\sigma$ uncertainties of the measured distances.  The colour-coding corresponds to the two sets of galaxies that we analyze (see Table \ref{tab:obj}).  Set B contains the outer LG galaxies (green points) as well as the MW and M31 (black).  Set A encompasses all galaxies shown in the figure, including the satellite systems of the MW (blue) and of M31 (red).}
   \label{fig:obj1}
   \end{center}
\end{figure}

\begin{figure*}
   \begin{center}
   \includegraphics[width=0.75\textwidth]{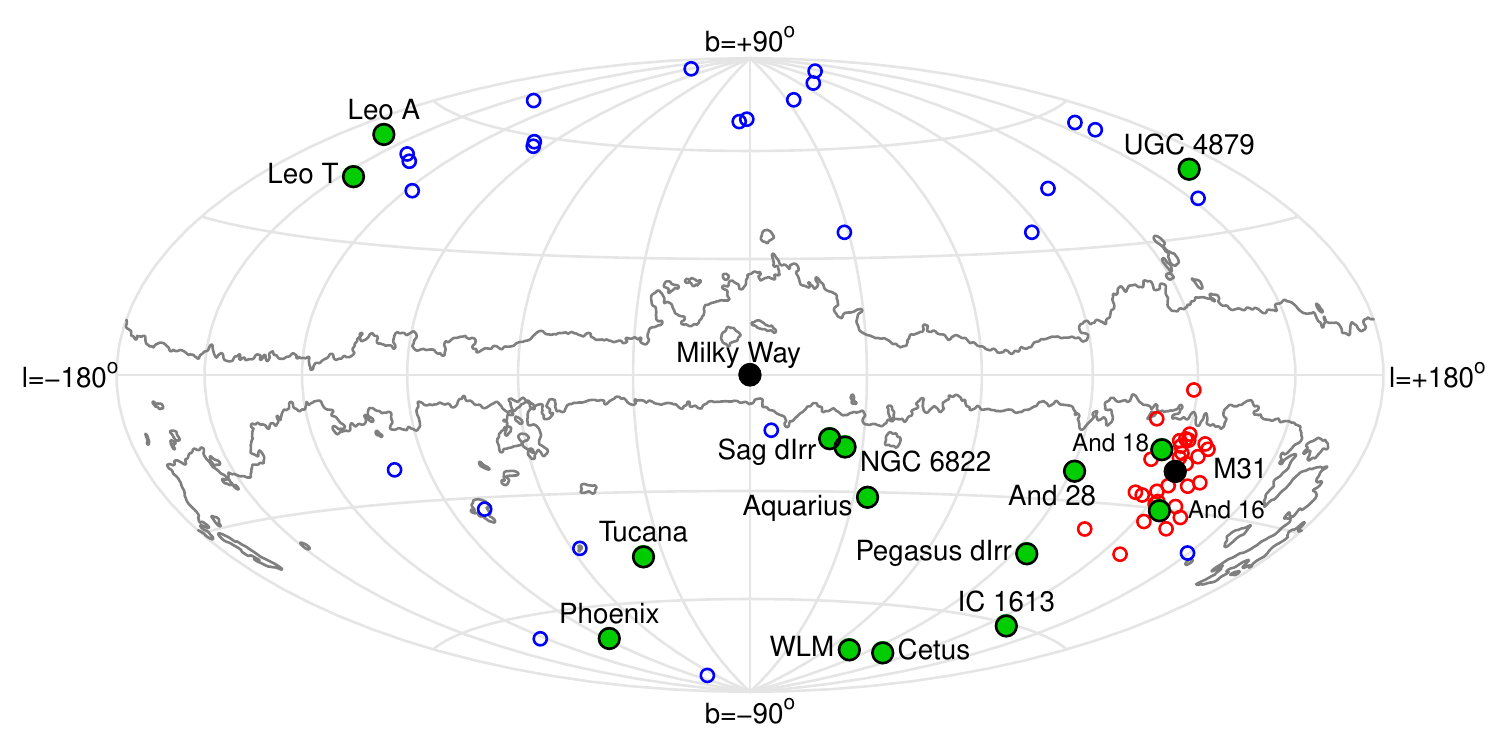}
   \caption{The on-sky distribution of the LG galaxies in galactic longitude $l$ and latitude $b$.  Only those galaxies depicted in Figure \ref{fig:obj1} are shown, and the same colour-coding is maintained.  The satellites of M31 (red) are clustered along the line-of-sight toward M31, whereas the satellites of the MW (blue) are scattered around the sky.  To explain the dearth of MW satellites at low galactic latitude, we draw contours of constant extinction $A(V)=1$ in the $V$-band from \citet{schlegel98}.}
   \label{fig:obj2}
   \end{center}
\end{figure*}

\subsubsection{The established method}
The challenge of determining the rest frame of the LG reduces to measuring $\vsunlg$, the solar motion with respect to the LG centre-of-mass.  This is because observable velocities are heliocentric and can be boosted to the LG rest frame once the Sun's motion in the LG is known.  There is a well-established method in the literature to measure this solar motion, although the statistical flavours vary (\citealt{yahil77, edlb82, kara96, rauzy, courteau99, tully2008}).  In this study we follow a Bayesian approach which requires us to define the likelihood function of our data and the prior distributions on our parameters.  With the underlying assumption that the radial velocities of LG members are Gaussian distributed, the likelihood function is
\begin{equation} \label{eq:like1d}
   \mathscr{L} = \prod_{i=1}^N \frac{1}{\sigma \sqrt{2\pi}} \exp \left(-\frac{(V_i + \los_i \cdot \bmath{V}_{\sun} )^2}{2\sigma^2} \right)
\end{equation}
where the product is taken over $N$ total LG galaxies, with object $i$ having an observed velocity $V_i$ along the line-of-sight unit vector $\los_i$, and where we have assumed the measured error on $V_i$ is negligible in comparison to $\sigma$.  There are four parameters in this function: the line-of-sight velocity dispersion $\sigma$, and the three components of $\vsunlg$, which is simply abbreviated as $\bmath{V}_{\sun}$ in (\ref{eq:like1d}).

The physics underlying the above likelihood function are simplistic but reasonable.  The LG is assumed to have a centre-of-mass which is moving relative to us (hence $\vsunlg$), and the galaxies within the LG are assumed to swarm randomly about this centre (hence $\sigma$).  This likelihood function would not apply if the velocities of the LG members were correlated in some way, e.g., if there were bulk rotation, flows of satellites, or accretion of groups onto the LG.  However, there is little observational motivation for adopting a physical model beyond the simple one given here.

\subsubsection{Excluding satellites}
An often overlooked aspect of equation (\ref{eq:like1d}) is that the data, i.e. the $N$ galaxies, are implicitly required to be \textit{independent} tracers of motion within the LG.  That is, if a galaxy's motion is governed not by the LG in bulk but rather by an individual LG member (namely, either the MW or M31) then it must be excluded from the product in (\ref{eq:like1d}).  Motivated perhaps by a lack of sufficient data, a number of previous authors have ignored this condition and included in their solution all known LG galaxies, including satellites (e.g. \citealt{kara96, courteau99, tully2008}; however, see \citealt{yahil77}).  

If we incorporate all satellites into the algorithm, the bulk motions of M31 and the MW will be over-represented, which will cause the outer members of the LG (i.e. the non-satellites) to receive smaller weight in the likelihood.  And because there are now more satellites known around M31 than the MW (\citealt{mcconn2012}), the algorithm would be biased toward the M31 rest frame.  In addition, the parameter $\sigma$ would have dubious physical meaning.

To account for these issues, we perform our analysis on two sets of LG galaxies summarized in Table \ref{tab:obj}: Set A, which comprises all 74 LG members within 1.5 Mpc of the MW and M31; and Set B, which excludes satellites by taking only the MW, M31, and those outlying members beyond 350 kpc of the MW and M31 (and within 1.5 Mpc).  As discussed above, Set A is a bad choice and will give biased results, but we include it in the analysis for the sake of comparison.  We take Set B as our adopted case since it plausibly contains independent tracers.  Figure \ref{fig:obj1} depicts the members of these sets as functions of distance from M31 and the MW, and Figure \ref{fig:obj2} shows their on-sky distribution in galactic coordinates.

We selected 1.5 Mpc as the outer boundary in order to exclude the four members of the Antlia subgroup ($D_{\rm MW} \approx 1.4$ Mpc, $D_{\rm M31} \approx 2.0$ Mpc), whose kinematics indicate likely participation in the Hubble flow (\citealt{vdb99ant}).  As such, we cannot consider these galaxies as members of the LG for the purposes of our study.  The lower bound of 350 kpc was chosen to provide good separation from the virial boundaries of the MW and M31, which are roughly 200 kpc (\citealt{shull2014}).  Figure \ref{fig:obj1} shows that the members of Set B are well separated from the clustered satellite populations of M31 and the MW.  Even though Andromeda 16 has a mean distance of only 324 kpc from M31, the error is large enough ($\pm$42 kpc) to place the galaxy beyond 350 kpc, and so we include it in Set B.

\subsubsection{Three-dimensional Likelihood}

In addition to including only independent tracers in our analysis, we improve upon previous studies by incorporating transverse motions into our likelihood function, where available.  The one-dimensional Gaussian of equation (\ref{eq:like1d}) is replaced with a three dimensional Gaussian of the form
\begin{equation} \label{eq:like3d}
   \mathscr{L} = \prod_{i=1}^N (2\pi)^{-3/2} |\bmath{\Sigma}_i|^{-1/2} \exp \left(-\frac{1}{2}\bmath{v}_i^{\rm T} \bmath{\Sigma}_i^{-1} \bmath{v}_i \right)
\end{equation}
where $\bmath{\Sigma}_i$ is a covariance matrix having determinant $|\bmath{\Sigma}_i|$, and $\bmath{v}_i$ is a composite vector (with transpose $\bmath{v}_i^{\rm T}$) of the form
\begin{equation}
   \bmath{v}_i = \left( \vwi - \east_i \cdot \bmath{V}_{\sun}, ~~\vni + \north_i \cdot \bmath{V}_{\sun}, ~~V_i + \los_i \cdot \bmath{V}_{\sun} \right)
\end{equation}
where the unit vectors pointing east, north, and along the line-of-sight for each object are $\east_i$, $\north_i$, and $\los_i$, respectively, and the measured velocities in these directions are $-\vwi$, $\vni$, and $V_i$.  Again, $\bmath{V}_{\sun}$ is shorthand for $\vsunlg$.

Because the errors on measured transverse velocities are generally non-negligible, $\bmath{\Sigma}_i$ is the sum of two matricies: one which characterizes the intrinsic dispersion of the LG, and one containing measurement errors.  Assuming that the intrinsic dispersion is isotropic, we use
\begin{equation}
\bmath{\Sigma}_i = \sigma^2 \bmath{I} + \bmath{\Sigma}_{ {\rm obs,} i},
\end{equation}
where $\sigma$ is the one-dimensional velocity dispersion as in equation (\ref{eq:like1d}), $\bmath{I}$ is the identity matrix, and $\bmath{\Sigma}_{ {\rm obs,} i}$ is the covariance matrix of the measured velocities for object $i$.  Our assumption that the velocities are Gaussian distributed requires that we treat the velocity dispersion $\sigma$ as a constant.  In other words, we do not allow $\sigma$ to be a function of distance within the LG, which formally implies an isothermal distribution for our tracers.

It would be a great benefit to our study if the proper motions of the outer LG galaxies were known, but unfortunately the only galaxies with measured transverse motions are the MW (\citealt{reid&brunthaler}), M31 (\citealt{sohn2012}), and some of their satellites (e.g. \citealt{sohn2013, nitya2013, vieira2010, piatek2008, brunthaler2007, brunthaler2005}).  Because Set A is intended only as a illustrative case, we do not take pains to include the proper motions of the satellite galaxies.  Thus, only the MW and M31 contribute measured transverse motions, and we adopt the values previously discussed in Section \ref{sec:tbg2}.  For objects lacking data in the transverse directions, we simply use the corresponding one-dimensional Gaussian in the likelihood product.

We adopt the parametrization $\vsunlg = \vsun \los_{\odot}$, where $\vsun$ is the amplitude and $\los_{\odot}$ is the direction of solar motion, specified according to equation (\ref{eq:los}) by a galactic longitude $\lsun$ and latitude $\bsun$.  This gives us four total parameters to measure: $\sigma$, $\vsun$, $\lsun$, and $\bsun$.  The priors on these parameters are chosen to be as uninformative as possible.  The angles $\lsun$ and $\bsun$ are given uniform priors on the sphere, and $\sigma$, being a scale parameter, is taken to be uniform in log space.  This choice for $\sigma$ is called the Jeffreys prior (\citealt{jeffreys46}), and we use lower and upper bounds of $10^{-2}$ and $10^3~\kms$, respectively.  Similarly, we impose the Jeffreys prior on the amplitude $\vsun$, with bounds of $10^{-2}$ and $10^3~\kms$.

We take the logarithm of equation (\ref{eq:like3d}) multiplied by our choice of priors as our log-posterior distribution, and we sample it with a standard Markov Chain Monte Carlo (MCMC) Metropolis-Hastings algorithm (\citealt{metropolis}).  We ensure convergence by running long chains ($10^6$ iterations), tossing away the first half of the chain as a burn-in phase, and thinning the chain by excluding every second entry.  Our final chains have a length of $2.5 \times 10^5$, and their convergence to smooth distributions is verified by inspecting the 2-d parameter correlations and corresponding 1-d marginalized histograms.  Acceptance rates for our chains fall within the desired range of 20\% to 30\%.

\subsection{Step 2: Measuring $\bmath{\mrat}$} \label{sec:meh2}

\subsubsection{Imposing momentum balance}

Stated in words, equation (\ref{eq:mom}) says that the respective momenta of M31 and the MW must be equal and opposite, which implies that $\vmwlg$ and $\vandlg$ must be antiparallel.  However, there is no guarantee that this condition will hold when we combine our solution for $\vsunlg$ with our chosen $\vmwsun$ and $\vandsun$ from Section \ref{sec:tbg2}.  In fact, as we will see in the next section, the measurement of $\vsunlg$ is sufficiently uncertain that the vectors $\vmwlg$ and $\vandlg$ can be oriented by a wide range of relative angles.  Therefore, our strategy is to \textit{impose} momentum balance using a Bayesian procedure, where the full covariance of all quantities will be taken into account.

Let us first summarize the results of the previous Step 1 of our method.  The posterior distributions for $\vmwlg$ and $\vandlg$ can be constructed from equation (\ref{eq:decomp}), where we take $\vsunlg = \vsun ( \cos \bsun \cos \lsun,~ \cos \bsun \sin \lsun,~ \sin \bsun )$ from our previous MCMC chain results, and where we randomly sample $\vmwsun$ and $\vandsun$ according to the parameter choices of Section \ref{sec:tbg2}.  As we will see in the next section, the components of $\vmwlg$ and $\vandlg$ are normally distributed.  Accordingly, the posterior distribution $P(\bmath{v})$ is summarized by a six-dimensional Gaussian of the form
\begin{equation} \label{eq:skpost}
   P(\bmath{v}) = (2\pi)^{-3/2} |\bmath{\Sigma}|^{-1/2} \exp \left(-\frac{1}{2}(\bmath{v}-\bmath{\bar{v}})^{\rm T} \bmath{\Sigma}^{-1} (\bmath{v}-\bmath{\bar{v}}) \right)
\end{equation}
where $\bmath{v}$ is a column vector of six entries formed by stacking the three components of $\vandlg$ below those of $\vmwlg$, and where $\bmath{\bar{v}}$ denotes the associated mean vector and $\bmath{\Sigma}$ the covariance matrix.

As it stands, equation (\ref{eq:skpost}) simply summarizes our previous results, but it can be the lynchpin which imposes momentum balance if we parametrize $\bmath{v}$ appropriately.  Let $\vmwlg$ point along the vector $\u$, specified by the galactic longitude $\lmw$ and latitude $\bmw$ as
\begin{equation} \label{eq:u}
   \u = ( \cos \bmw \cos \lmw ,~ \cos \bmw \sin \lmw ,~ \sin \bmw ).
\end{equation}
Now we require $\vandlg$ to point in the opposite direction, toward longitude $\lmw+180^{\circ}$ and latitude $-\bmw$, which is simply the unit vector $-\u$.  Denoting the magnitude of the velocities as $\vmw$ and $\vand$, equation (\ref{eq:mom}) now takes the symmetric form
\begin{equation} \label{eq:momu}
   0 = \u ( M_{\rm MW} \vmw - M_{\rm M31} \vand ).
\end{equation}
With this parametrization, the components of $\bmath{v}$ are $\vmwlg = \vmw \u$ and $\vandlg = -\vand \u$, or explicitly,
\begin{eqnarray}
   \bmath{v} = ( \vmw \cos \bmw \cos \lmw ,~ \vmw \cos \bmw \sin \lmw , \nonumber \\
	\vmw \sin \bmw ,~ -\vand \cos \bmw \cos \lmw , \nonumber \\
	-\vand \cos \bmw \sin \lmw ,~ -\vand \sin \bmw ).
\end{eqnarray}
This choice for $\bmath{v}$ transforms equation (\ref{eq:skpost}) into a likelihood function of the data given the parameters $\vand$, $\vmw$, $\lmw$, and $\bmw$.  Consequently, estimating these parameters will effectively ``twist" the initial values of $\vmwlg$ and $\vandlg$ into anti-alignment.  Once this is done, it will be straightforward to measure the mass ratio from equation (\ref{eq:momu}) as $\mrat = \vmw / \vand$.

The prior distributions for $\lmw$ and $\bmw$ are chosen to be uniform on the sphere, and the priors for $\vmw$ and $\vand$ are chosen to be lognormal with parameters $\sigma=1/\sqrt{2}$ and $\mu=4$.  Because the mass ratio is evaluated as $m=\vmw / \vand$, these choices imply that the prior on $\log_{10} m$ is normally distributed, is centered on 0, and has a standard deviation of 1.  This is a sufficiently broad prior, as it places 95.5\% confidence (i.e. 2$\sigma$ limit) that neither galaxy is more than 100 times as massive as the other.  Importantly, these choices allow M31 and the MW to be treated symmetrically.\footnote{For instance, symmetry requires the prior on $m=\mrat$ to have the same probability in the range 0.1 to 1 as in the range 1 to 10.  From this, it is clear that our lognormal prior on $m$ is justifiable.}

As in Section \ref{sec:meh1}, we multiply the likelihood by our chosen priors, and we sample the logarithm of this product by a standard MCMC Metropolis-Hastings algorithm.  The resulting chains are processed in the same way, and convergence is checked in the same way.  We get acceptance rates that once again fall in the desired range of 20\% to 30\%.

\subsubsection{Updated inferences} \label{sec:twist}

After running the MCMC chain and retrieving our parameter estimates, our main interest is evaluating $\mrat = \vmw / \vand$.  But as a bonus we also get a new posterior inference on $\vsunlg$.  From equation (\ref{eq:decomp}) we find
\begin{equation} \label{eq:vsunlg.twist}
   \vsunlg =  \vmw \u - \vmwsun, 
\end{equation}
where $\u$ is given by equation (\ref{eq:u}) in terms of $\lmw$ and $\bmw$.  The values for $\vmw$, $\lmw$, and $\bmw$ are provided by their MCMC chain values, and $\vmwsun$ is randomly sampled from Gaussian probabilities and the parameter choices of Section \ref{sec:tbg2}.  This inference on $\vsunlg$ is now consistent with the balance of momentum within the LG, and can be compared with the original inference (Step 1, Section \ref{sec:meh1}) to gauge the amount of ``twisting" that occurred.

We can also test the degree of twisting in the tangential motion of M31 by determining the posterior inferences on $\vn$ and $\vw$.  From equation (\ref{eq:decomp}) and (\ref{eq:vm31.wrt.sun}) we find
\begin{equation} \label{eq:vnvw.twist}
   \vw \east - \vn \north  =  \vlos \los + \vsunlg + \vand \u,
\end{equation}
where $\u$ is given by equation (\ref{eq:u}) in terms of $\lmw$ and $\bmw$, and $\vsunlg$ is taken from equation (\ref{eq:vsunlg.twist}).  Once again, the MCMC chain determines the values for $\vand$, $\lmw$, and $\bmw$.  Because $\east$, $\north$, and $\los$ are mutually orthogonal unit vectors, the posterior for $\vw$ is found by dotting the above equation with $\east$, and that of $\vn$ is found by dotting with $-\north$.

\section{MCMC Results} \label{sec:res}

\subsection{Step 1: $\bmath{\vsunlg}$} \label{sec:res1}

\begin{table*}
   \begin{center}
   \caption{Measurement of $\vsunlg$ from the MCMC results of Step 1 (first two rows) and of Step 2 (final two rows).  The parametrization defined in Section \ref{sec:meh} gives $\vsunlg$ in terms of an amplitude $\vsun$ and a direction determined by the galactic longitude $\lsun$ and latitude $\bsun$.  These quantities are directly measured in Step 1, but in Step 2 we must compute them using equation (\ref{eq:vsunlg.twist}).  We also give the projections of $\vsunlg$ along the unit vectors $\rhat$, $\that$, and $\phat$ in the final three columns.  The parameter $\sigma$ is the scatter about the fit in Step 1 and is not computed in Step 2.  The quoted values of each quantity are the mean and standard deviation.}
   \begin{tabular}{lccccccccc}
   \hline
   &&        $\sigma$ & $\vsun$ & $\lsun$ & $\bsun$ && $\rhat \cdot \vsunlg$ & $\that \cdot \vsunlg $ & $\phat \cdot \vsunlg$ \\
   &&        (km s$^{-1}$) & (km s$^{-1}$) & (deg) & (deg) && (km s$^{-1}$) & (km s$^{-1}$) & (km s$^{-1}$) \\
   \hline
   \multicolumn{10}{l}{Step 1: Intermediate results} \\
   ~~Set A  && $100.4 \pm 8.3$ & $307.2 \pm 18.3$ & $102.5 \pm 5.4$ & $-8.4 \pm 3.8$ && $281.3 \pm 15.8$ & $-63.0 \pm 21.1$ & $-100.1 \pm 29.6$ \\
   ~~Set B  && $58.1 \pm 10.5$ & $312.1 \pm 24.6$ & $103.5 \pm 4.9$ & $-8.1 \pm 4.3$ && $287.4 \pm 23.3$ & $-66.1 \pm 23.8$ & $-96.2 \pm 26.6$ \\
   \multicolumn{10}{l}{Step 2: Final results} \\
   ~~Set A           && $-$ & $302.4 \pm 13.5$ & $98.3 \pm 3.6$ & $-6.8 \pm 2.9$ && $268.6 \pm 12.6$ & $-67.4 \pm 15.2$ & $-119.0 \pm 19.1$ \\
   ~~Set B$^\dagger$ && $-$ & $298.8 \pm 14.7$ & $98.4 \pm 3.6$ & $-5.9 \pm 3.0$ && $264.3 \pm 15.5$ & $-70.8 \pm 15.4$ & $-117.3 \pm 18.1$ \\
   \hline
   \multicolumn{10}{l}{$^\dagger$Denotes the adopted case.}
   \label{tab:res1}
   \end{tabular}
   \end{center}
\end{table*}

\begin{figure}
   \begin{center}
   \includegraphics[width=0.45\textwidth]{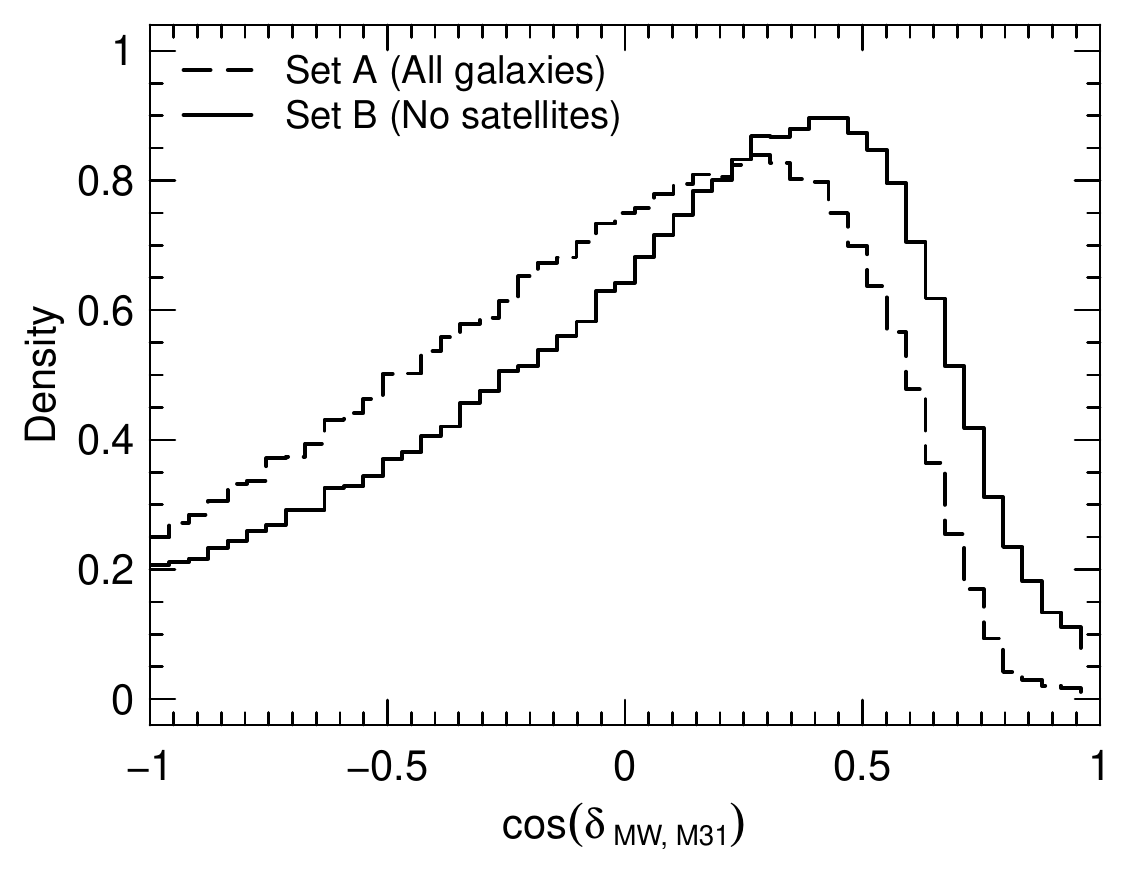}
   \caption{Probability distributions of $\cos \angmm$ derived from the MCMC results of Step 1 for Set A (dashed) and Set B (solid).  The angle $\angmm$ is the angle between the velocity vectors $\vmwlg$ and $\vandlg$, which are constructed as in equation (\ref{eq:decomp}).  Strict momentum balance requires $\cos \angmm = -1$ as is clear from equation (\ref{eq:mom}), but our derived distributions are sufficiently broad to allow a wide range of orientations for the vectors $\vmwlg$ and $\vandlg$.  This is essentially due to the uncertainty in the measured quantity $\vsunlg$.}
   \label{fig:cosang}
   \end{center}
\end{figure}

Here we implement Step 1 of our method, as given in Section \ref{sec:meh1}.

The first two rows of Table \ref{tab:res1} summarize the results of our MCMC chains, giving the mean and standard deviation for each parameter.  Within their respective errors, the values for the solar motion parameters $\vsun$, $\lsun$, and $\bsun$ are consistent between Sets A and B.  However, $\sigma$ changes dramatically, with the value for Set A being nearly twice as large as the value for Set B.  This is not terribly surprising because Set A includes satellites, which causes $\sigma$ to encompass not only the random motions pertaining to the LG but also the random motions within the halos of M31 and the MW.  We will return to the physical significance of $\sigma$ in Section \ref{sec:lgmass}, where we will use it in our estimate of the dynamical mass of the LG.

The relative sizes of the errors (i.e. standard deviations) for our parameters generally increase from set A to B as seen in Table \ref{tab:res1}.  This is almost certainly tied to the fewer number of objects in Set B (Table \ref{tab:obj}) and the sparser sky coverage (Figure \ref{fig:obj2}).  The magnitude of this effect is small, however, and it is remarkable that the absolute size of the errors do not markedly increase between sets A and B, despite roughly a factor of four difference in the total number of objects.

The projected components of $\vsunlg$ as measured along the $\rhat$, $\that$, and $\phat$ directions are given in the final three columns of \ref{tab:res1}.  It is worth noting that the dominant component of the vector $\vsunlg$ is along $\rhat$, with the mean value for Set B ($\sim$$290~\kms$) more than double the amplitude in the tangential directions ($\sim$$120~\kms$, taking the quadrature sum of the means).

Having measured $\vsunlg$, we are now in a position to construct the velocities $\vmwlg$ and $\vandlg$ using equation (\ref{eq:decomp}) and the parameter choices of Section \ref{sec:tbg2}.  Of principal interest is whether or not these vectors point in opposite directions as required by the condition of momentum balance in equation (\ref{eq:mom}).  Letting $\angmm$ denote the angle between $\vmwlg$ and $\vandlg$, Figure \ref{fig:cosang} gives the probability distributions of $\cos \angmm$ for Set A and B.  It is clear that anti-alignment of the vectors, though possible, is far from preferred.  In fact, the uncertainties are so large and the range of possible orientations is so broad that even mutual orthogonality is quite likely!

At this juncture, one might be tempted to raise a red flag and claim that equation (\ref{eq:mom}) does not hold.  However, such a claim can be justified only if the distributions of Figure \ref{fig:cosang} have \textit{negligible} probability at $\cos \angmm \approx -1$.  This is not the case.  We accordingly interpret Figure \ref{fig:cosang} as a reflection of the probabilistic nature of the problem and the large uncertainties on the measurement of $\vsunlg$.  Given that our goal is to measure $\mrat$, Figure \ref{fig:cosang} also emphasizes the need to impose momentum balance in the following section.

\subsection{Step 2: $\bmath{\mrat}$ and updated inferences} \label{sec:res2}

\begin{table*}
   \begin{center}
   \caption{Summary of results after imposing momentum balance in Step 2.  The quantities which are directly measured in the MCMC chains are $\vand$, $\vmw$, $\lmw$, and $\bmw$ which are defined in Section \ref{sec:meh2}.  These quantities can be combined to yield posterior estimates of the mass ratio $\mrat = \vmw / \vand$.  As discussed in the text, the logarithm of the mass ratio is distributed normally whereas the mass ratio itself is not.  Consequently, we list $\log_{10} (\mrat)$ in the first column.  In addition, posterior estimates of the M31 (heliocentric) transverse velocities $\vn$ and $\vw$ can be computed according to equation (\ref{eq:vnvw.twist}) and are listed in the final two columns.  The quoted values of each quantity are their mean and standard deviation.}
   \begin{tabular}{lccccccc}
   \hline
   & $\log_{10} (\mrat)$ & $\vand$ & $\vmw$ & $\lmw$ & $\bmw$  & $\vn$ & $\vw$ \\
   & --- & (km s$^{-1}$) & (km s$^{-1}$) & (deg) & (deg) & (km s$^{-1}$) & (km s$^{-1}$) \\
   \hline
   A       & ~~$0.44 \pm 0.24$ & $33.2 \pm 12.8$ & $85.7 \pm 13.9$ & $139.5 \pm 14.1$ & $-30.4 \pm 10.0$ & $-58.3 \pm 21.2$ & $-109.7 \pm 26.0$ \\
   B$^\dagger$& ~~$0.36 \pm 0.29$ & $38.1 \pm 16.3$ & $81.3\pm 17.2$ & $141.6\pm 13.5$ & $-28.3\pm 10.4$ & $-62.1\pm 22.5$ & $-105.7 \pm 25.4$ \\
   \hline
   \multicolumn{7}{l}{$^\dagger$Denotes the adopted case.}
   \label{tab:res2}
   \end{tabular}
   \end{center}
\end{table*}

\begin{figure}
   \begin{center}
   \includegraphics[width=0.45\textwidth]{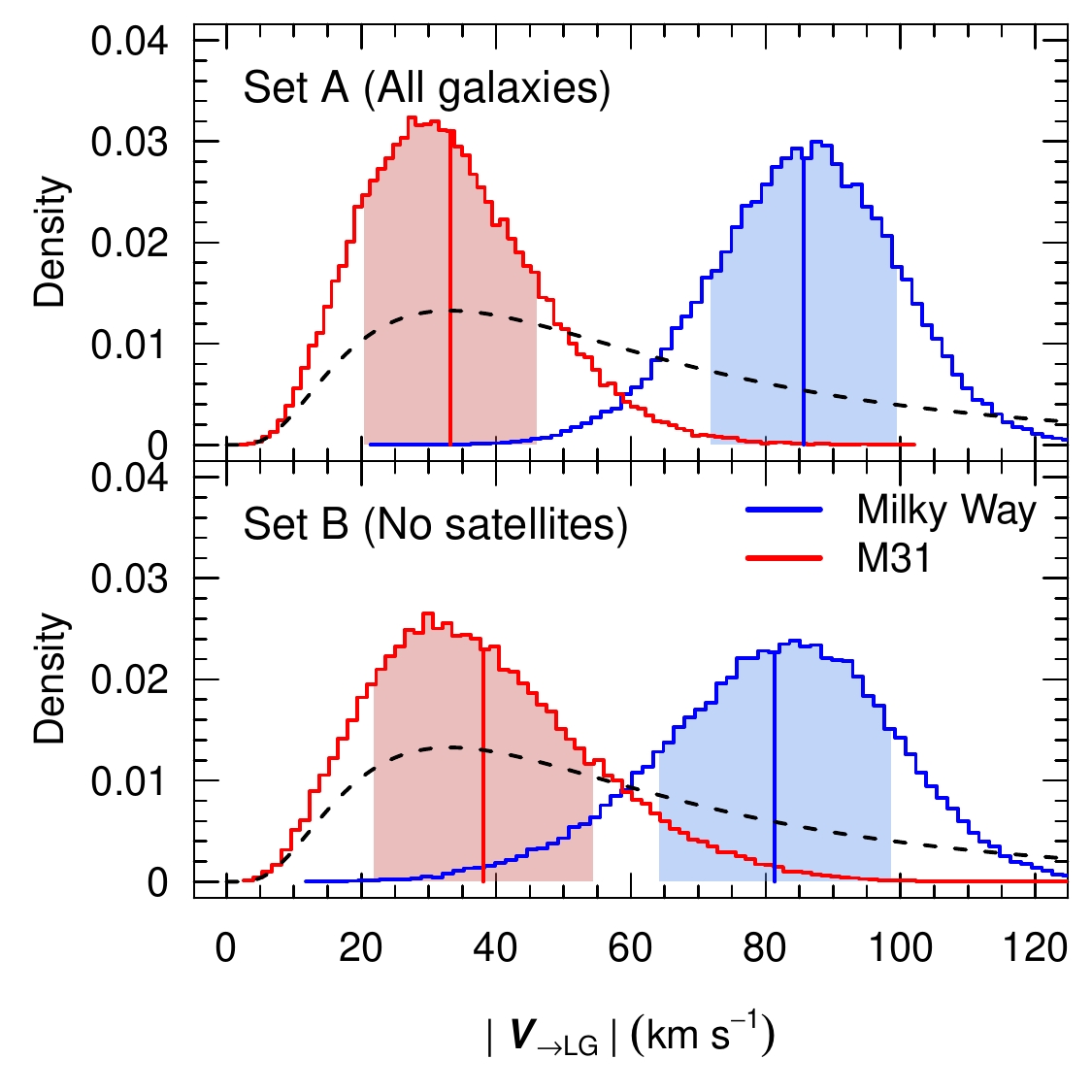}
   \caption{Probability distributions for the velocities $\vmw$ (blue) and $\vand$ (red) with respect to the LG, derived from the MCMC results of Step 2 for (top) Set A and (bottom) Set B.  The velocities shown here are properly anti-aligned according to equation (\ref{eq:momu}).  Due to the choice of lognormal priors (shown as dotted lines), the distributions exhibit slight skewness.  The mean values are given by vertical lines and $1\sigma$ uncertainties are shaded.}
   \label{fig:vmvm2}
   \end{center}
\end{figure}

\begin{figure}
   \begin{center}
   \includegraphics[width=0.45\textwidth]{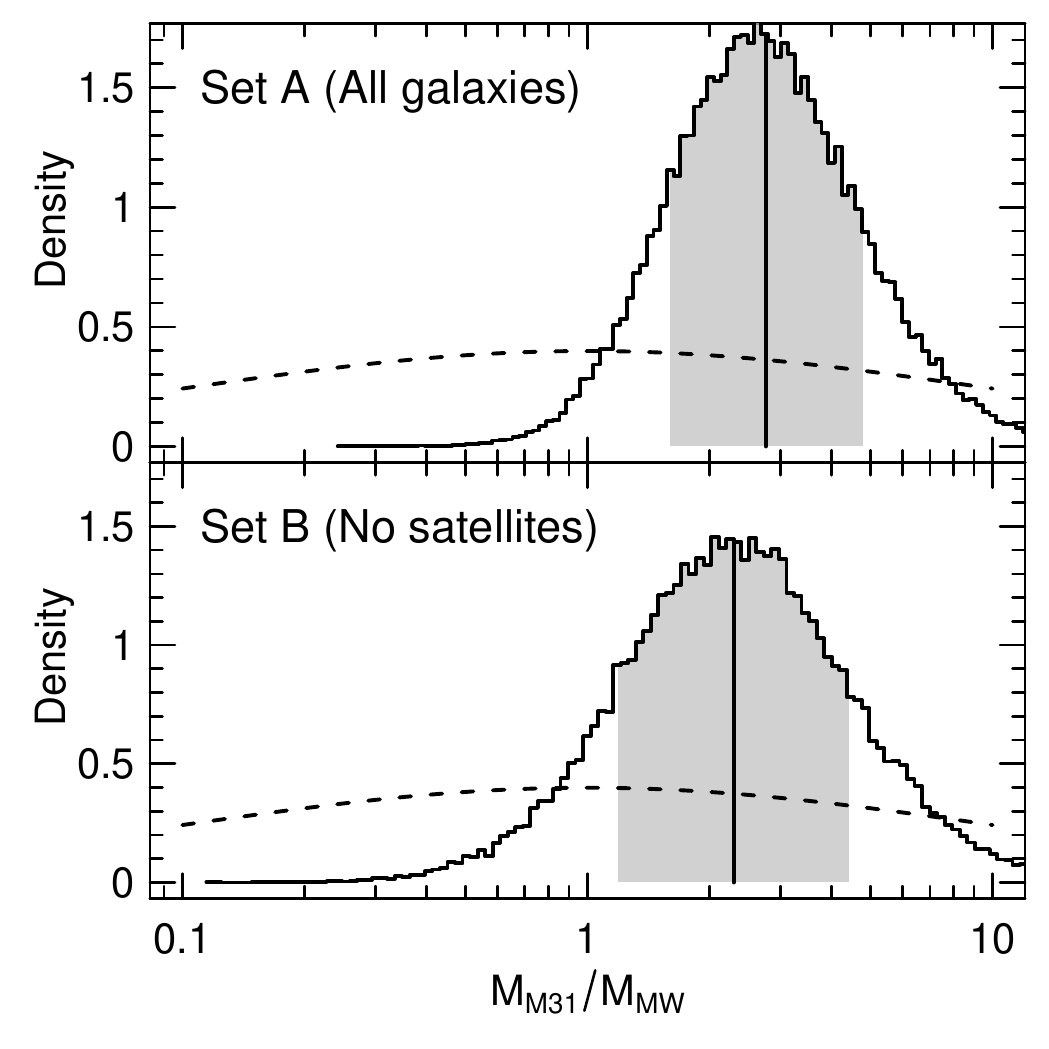}
   \caption{Posterior inference on the mass ratio of M31 and the MW, computed from the MCMC results of Step 2 as $\mrat= \vmw / \vand$ for (top) Set A and (bottom) Set B.  The assumed prior (dotted line) is relatively flat in comparison to the posterior distributions.  The mean values are given by solid lines, and the $1\sigma$ uncertainties are shown as shaded regions.}
   \label{fig:mrat}
   \end{center}
\end{figure}

\begin{figure*}
   \begin{center}
   \includegraphics[width=0.9\textwidth]{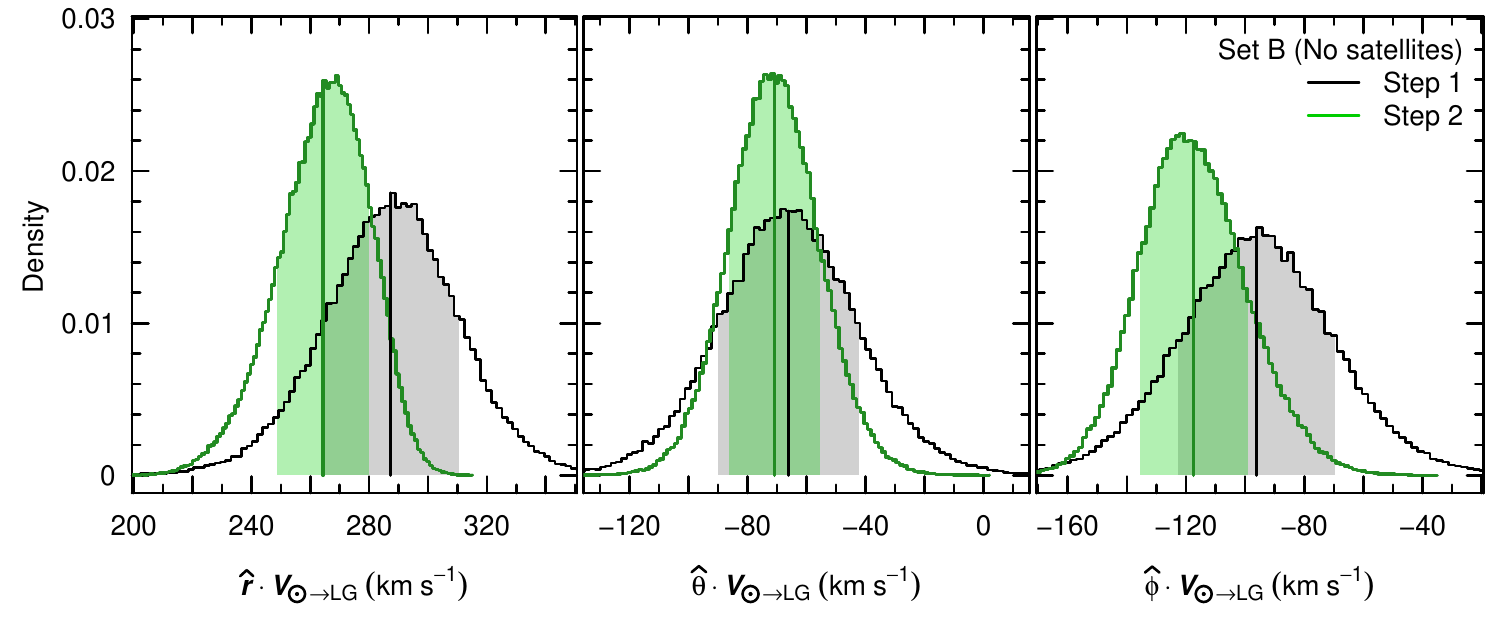}
   \caption{Probability distributions for $\vsunlg$ from the MCMC results of the adopted Set B after Step 2 (green) as compared to Step 1 (black).  We compute the distributions for Step 2 using equation (\ref{eq:vsunlg.twist}). The velocities are projected into components along (left) $\rhat$, (middle) $\that$, and (right) $\phat$, which allows a direct comparison of the ``twisting" induced by Step 2 of our method.  The mean value of each component shifts leftward, the $1\sigma$ uncertainties (shaded regions) decrease in size, and the final distributions have a slight skewness.  The distributions for Step 1 and 2 correspond to the intermediate and final inferences, respectively, listed in Table \ref{tab:res1} for Set B.}
   \label{fig:vsunB}
   \end{center}
\end{figure*}

\begin{figure}
   \begin{center}
   \includegraphics[width=0.45\textwidth]{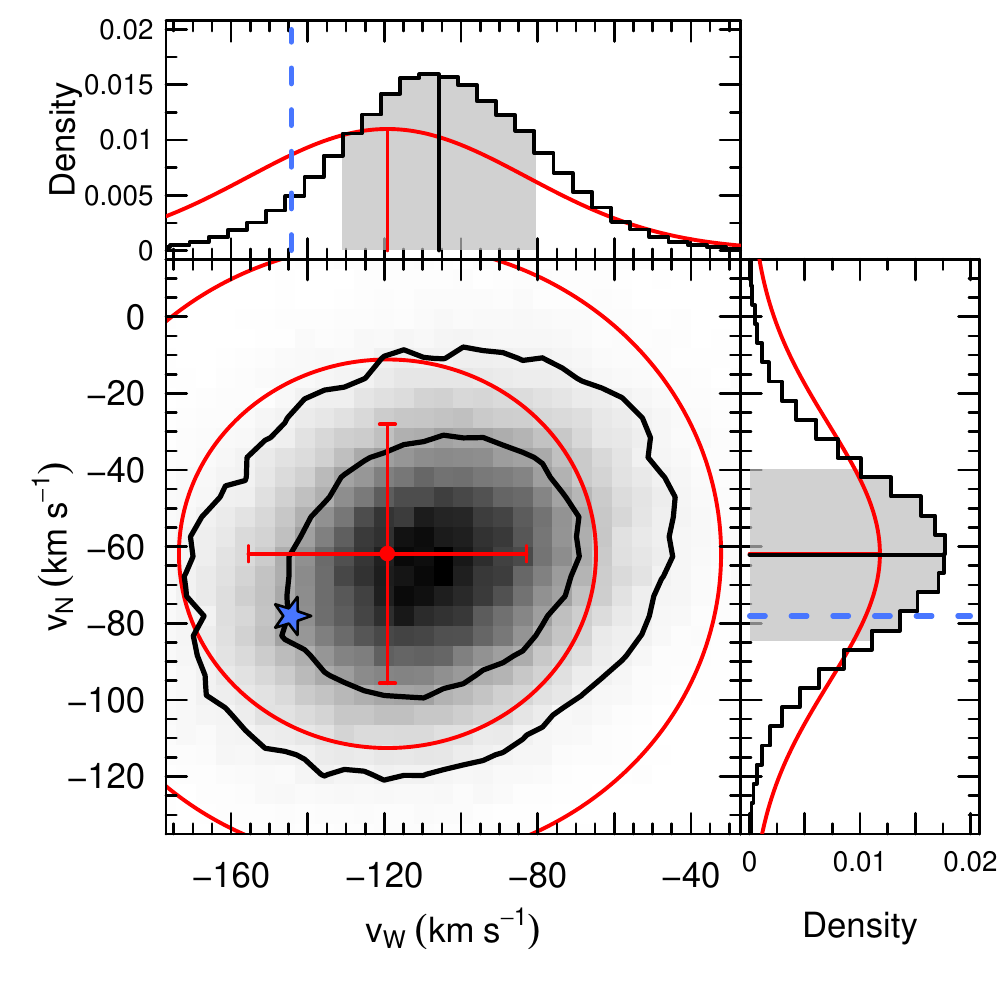}
   \caption{2-d and 1-d probability distributions for $\vn$ and $\vw$ derived from the MCMC results of Set B Step 2 using equation (\ref{eq:vnvw.twist}).  For the 1-d marginalized distributions of $\vw$ (top) and $\vn$ (right), the shaded region depicts the $1\sigma$ uncertainty and the solid black line gives the mean value.  In the 2-d distribution, the black contours are drawn to contain 68.3\% and 95.5\% of the total probability in the plane, and the shading corresponds to an arbitrarily normalized density.  The values of $\vn$ and $\vw$ corresponding to a radial MW-M31 orbit are shown as dashed blue lines in the top and right panels, and as a blue star in the 2-d plane.  Also shown in red are the original parameter values for $\vn$ and $\vw$ as assumed in Section \ref{sec:tbg2}, where the top and right panels give the Gaussian curves and their mean values, and in the 2-d plane a red cross gives the mean and the $1\sigma$ error bars.  The red ellipses give the corresponding 1- and $2\sigma$ contours containing 68.3\% and 95.5\% of the total probability in the plane, respectively.  Note that these ellipses enclose a larger region than the formal $1\sigma$ error bars, which contain only 39.5\% of the probability in two dimensions.}
   \label{fig:cov-pm}
   \end{center}
\end{figure}

Here we implement Step 2 of our method, detailed in Section \ref{sec:meh2} and highlighted as follows: we use a Bayesian procedure to twist the vectors $\vmwlg$ and $\vandlg$ into anti-alignment, as conditioned by the balance of momentum in the LG (equations (\ref{eq:mom}) and (\ref{eq:momu})).  Consequently, we are able to derive the probability distribution for the mass ratio $\mrat$, as well as an updated posterior distribution for $\vsunlg$.

Table \ref{tab:res2} provides the results of our MCMC chains, summarized as the mean and standard deviation for the parameters $\vand$, $\vmw$, $\lmw$, and $\bmw$.  As in Section \ref{sec:res1}, the parameter values of Set A have generally smaller uncertainties than those of Set B.  The values of $\lmw$ and $\bmw$, which parametrize the direction of $\vmwlg$ according to equation (\ref{eq:u}), are consistent with the radial direction toward M31.  Projecting $\rhat$ as given in equation (\ref{eq:rpt}) onto the sky gives $l=121.7^{\circ}$ and $b=-21.5^{\circ}$.  These values are within the $2\sigma$ and $1\sigma$ errors of $\lmw$ and $\bmw$, respectively, as listed in Table \ref{tab:res2}.

The probability distributions for $\vmw$ and $\vand$ are depicted in Figure \ref{fig:vmvm2}.  Some of these distributions possess a slight skewness, which derives from the chosen (lognormal) prior.  The vectors $\vmwlg$ and $\vandlg$ are now properly anti-aligned, which means it is now straightforward to measure the mass ratio $\mrat$ by dividing our MCMC chain values for $\vmw$ by those of $\vand$.  The resulting probability distributions are shown in Figure \ref{fig:mrat}.  As required, the prior (dashed line) is uninformative with respect to the posterior distributions.  Because $\mrat$ is distributed normally in log space, we summarize our results in Table \ref{tab:res2} as the mean and standard deviation of $\log_{10} (\mrat)$.

Weighted by the many satellites of M31, it is unsurprising that the analysis of Set A favours M31 to be heavier than the MW.  Set B excludes these satellites and therefore provides more believeable results, although the final results are not significantly different to those of Set A.  The inference for Set B yields a mean value of $\mrat=2.30$ whereas Set A yields 2.76.  As Figure \ref{fig:mrat} shows, the uncertainty is large enough to allow a small probability (9.8\% for Set B) that the MW is more massive.

As outlined in Section \ref{sec:twist}, we can derive an updated posterior estimate of $\vsunlg$ from the MCMC chains of Step 2.  We can think of the vector as ``twisting" in response to the imposed condition of momentum balance.  The last three columns of Table \ref{tab:res1} give our final inferences as parametrized by the amplitude $\vsun$ and the angles $\lsun$ and $\bsun$.  As compared to our intermediate results (Table \ref{tab:res1}, Step 1), the parameters have shifted to new values and their uncertainties have become smaller.

Figure \ref{fig:vsunB} compares the $\rhat$, $\that$, and $\phat$ components of our final inference on $\vsunlg$ from Set B (green) to the corresponding intermediate inference (black).  It is apparent that the twisting imparts a noticeable change in the mean values of each component (up to $23~\kms$ different), but nevertheless the new values are roughly consistent with the intermediate values within their 1$\sigma$ errors.  The final posterior distributions also have a slight skewness, but this is only a minor effect.  These results confirm previous expectations (\citealt{edlb82}) that momentum balance can effectively constrain the components of $\vsunlg$.

Section \ref{sec:twist} also shows how we may derive posterior estimates for the transverse (heliocentric) motion of M31.  Figure \ref{fig:cov-pm} plots the 2-d posterior of $\vn$ and $\vw$ along with their corresponding marginalized 1-d distributions from Set B.  Also shown is our original assumption for the M31 transverse motion (red ellipses and lines; Section \ref{sec:tbg2}) and the values of $\vn$ and $\vw$ pertaining to a radial MW-M31 orbit (blue star and dashed lines).  By inspecting the 1$\sigma$ contours (enclosing 68.3\% of the probability in the 2-d plane), we see that the radial orbit just barely fits within the formal uncertainty of the final posterior distribution.  This is largely because the uncertainties decrease by $\sim$$10~\kms$ after we impose momentum balance.  Figure $\ref{fig:cov-pm}$ also shows that the updated mean values of $\vn$ and $\vw$ differ only marginally from our original values ($<1~\kms$ and $15~\kms$, respectively).  This conclusion is true for both Set A and B (Table \ref{tab:res2}).

From Figures \ref{fig:vsunB} and \ref{fig:cov-pm} we conclude that the condition of momentum balance has a stronger effect on $\vsunlg$ than on the transverse motion of M31 (i.e. it suffers more ``twisting").  This must be related to the fact that $\vsunlg$ appears twice in equation (\ref{eq:mom}) owing to the vector decompositions of equation (\ref{eq:decomp}).


\section{The Local Group Rest Frame} \label{sec:lgrf}

\begin{figure}
   \begin{center}
   \includegraphics[width=0.45\textwidth]{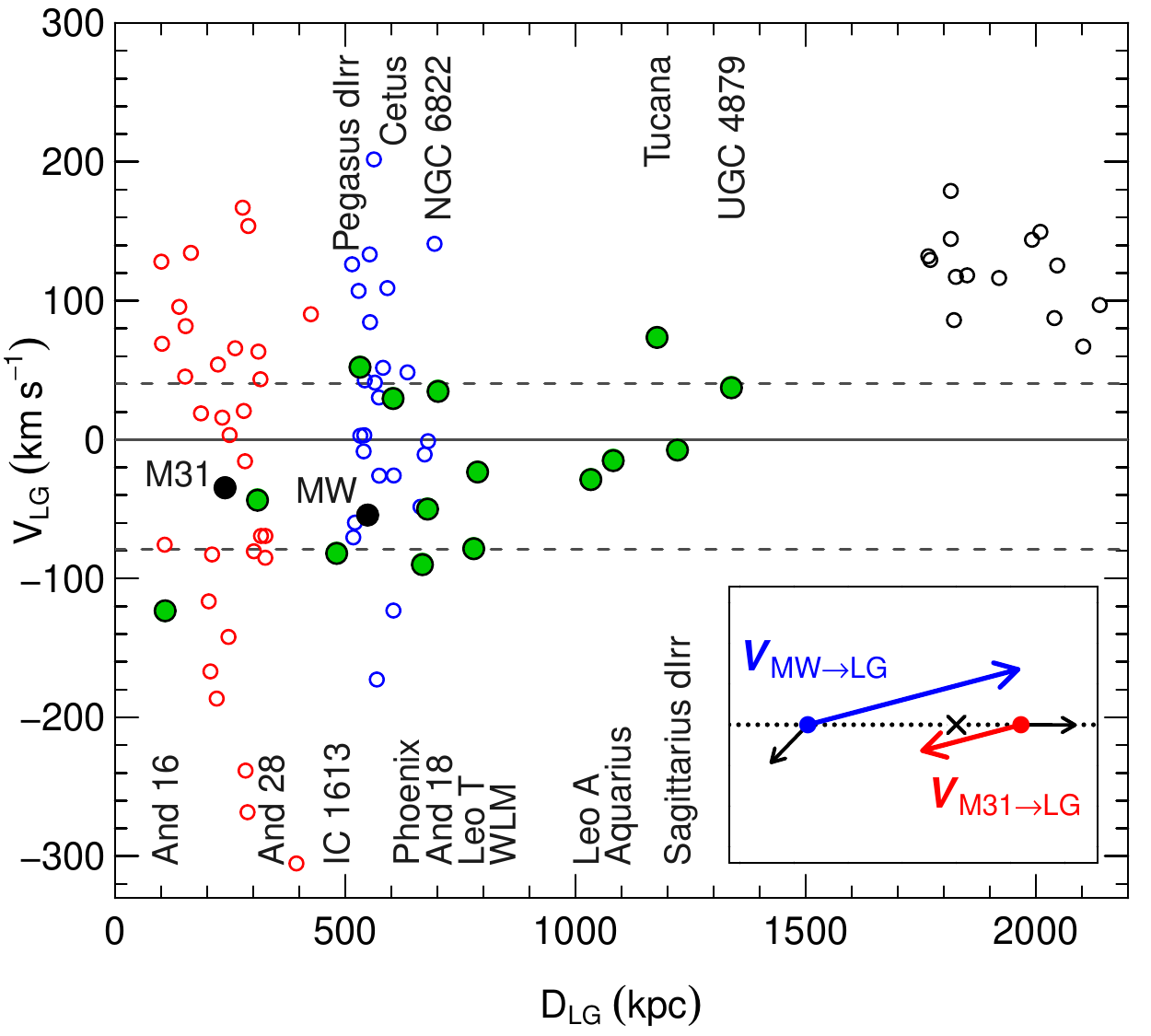}
   \caption{Distance $D_{\rm LG}$ to the LG centre-of-mass plotted against the line-of-sight velocity $V_{\rm LG}$ corrected to the LG rest frame.  These quantities are calculated as described in Section \ref{sec:lgrf}. The LG galaxies are the same as those of Figures \ref{fig:obj1} and \ref{fig:obj2}, and the same colour-coding has been maintained.  The names of the galaxies in Set B with velocities $V_{\rm LG}>0$ are placed at the top of the figure, and those with $V_{\rm LG}<0$ are placed at the bottom.  Additionally, we show the galaxies of the \citet{mcconn2012} catalogue which neighbour the LG and have velocities consistent with the Hubble flow ($D_{\rm LG}>1.5$~Mpc, small black circles).  The dashed lines indicate the value of the velocity dispersion ($59.7~\kms$) used in Section \ref{sec:lgmass} for the measurement of total mass within the LG. Inset (bottom right): Schematic showing the relative positions and velocities of the MW and M31 with respect to the LG centre (black cross).  The velocity vectors $\vmwlg$ (blue arrow) and $\vandlg$ (red arrow) are scaled arbitrarily, but their relative magnitude and direction reflect our final results from Section \ref{sec:res2}.  The small black arrows represent the line-of-sight unit vectors $\los$ toward each galaxy.  For M31, the line-of-sight direction is coincident with the radial direction (dotted line). }
   \label{fig:obj3}
   \end{center}
\end{figure}

Here, we use the results of Set B to calculate positions and velocities relative to the LG rest frame.  The position vector $\bmath{R}_{\rm LG}$ of the LG centre-of-mass is
\begin{equation} \label{eq:rlg}
   \bmath{R}_{\rm LG} = \bmath{R}_{\rm MW} + \frac{m}{1+m} \left( \bmath{R}_{\rm M31} - \bmath{R}_{\rm MW} \right),
\end{equation}
where $\bmath{R}_{\rm MW}$ and $\bmath{R}_{\rm M31}$ are position vectors of the MW and M31, respectively, and $m=\mrat$ takes the value of 2.30, which is the mean for Set B as derived from Table \ref{tab:res2}.  It is clear from this vector sum that the centre-of-mass lies on the line joining the MW and M31, with the location weighted by the total fraction of mass in each halo.  Taking the observed (heliocentric) distances of LG objects from \citet{mcconn2012}, we can now compute the position vector of each object relative to the LG centre-of-mass.

Because heliocentric transverse motions are not measured for many LG members, the full velocity vectors in the LG rest frame cannot be constructed.  Nevertheless, one can ``correct" the heliocentric line-of-sight velocity of any object to the LG rest frame by adding the projection of the solar motion:
\begin{eqnarray} \label{eq:vlosc}
   V_{\rm LG} & = & \vlos + \los \cdot \vsunlg \nonumber \\
   & = & \vlos -43 \cos b \cos l +293 \cos b \sin l -31  \sin b,
\end{eqnarray}
where $\vlos$ is the observed (heliocentric) line-of-sight velocity, $\los$ is the line-of-sight given by the object's galactic longitude $l$ and latitude $b$ in equation (\ref{eq:los}), and $\vsunlg$ is parametrized by our mean results for Set B (Table \ref{tab:res1}) as shown.

These corrected velocities can be difficult to interpret, because the projected direction $\los$ may have little relevance to the frame of interest.  In addition, if the objects within a set are scattered on the sky (as are the LG members), their projections along respective line-of-sight vectors will probe many different directions.  With these caveats in mind, we take the \citet{mcconn2012} catalog and compute the corrected velocities using (\ref{eq:vlosc}), which we show in Figure \ref{fig:obj3} as a function of the corresponding distance from the LG centre-of-mass.

A visual inspection of Figure \ref{fig:obj3} suggests that the velocity dispersion $\sigma$ of the outer LG galaxies (green dots) is reasonably well characterized by a single average value.  There is, however, possible evidence for $\sigma$ varying as a function of distance in the LG.  Splitting our tracers into an inner ($D_{\rm LG}$$<$750 kpc) and outer ($>$750 kpc) population, we find velocity dispersions of $66.2 \pm 4.7 ~\kms$ and $51.3 \pm 4.8 ~\kms$, respectively.  These values are each within two standard deviations of the velocity dispersion of the entire population, which takes the value $59.7 \pm 4.1 ~\kms$.

The inset of Figure \ref{fig:obj3} provides a schematic of the relative positions and velocities of the MW and M31 in the LG frame.  As shown, both galaxies are infalling toward the LG centre, and their velocity vectors are antiparallel by construction since $\vmwlg = \vmw \u$ and $\vandlg = -\vand \u$.  The velocities point in a direction which is somewhat offset from the radial direction in the LG (dotted line in the figure).  Taking $\u$ to orient the velocities as in equations (\ref{eq:u}) and (\ref{eq:momu}), and taking $\rhat$ to point along the radial direction per equation (\ref{eq:rpt}), we calculate $\u \cdot \rhat = 0.91 \pm 0.07$.  In other words, there is a small angle of roughly $23^{\circ}$ between $\u$ and $\rhat$ as shown in the schematic, but the uncertainty is large enough for the velocities to be consistent with radial at the 1.3$\sigma$ level.

As viewed from the LG centre, the radial direction toward the MW is $-\rhat$ and that toward M31 is $\rhat$.  Consequently, the projected radial velocities of the MW and M31 are both negative in the LG rest frame, denoting infall.  Likewise, the projected line-of-sight velocities are also negative, where the direction for $\los$ pertaining to each galaxy is indicated by a small black arrow in the schematic inset.  This is echoed in the main panel of Figure \ref{fig:obj3}, where the values for $V_{\rm LG}$ fall below zero for both the MW and M31.

A key assumption underpinning our analysis is that the LG in its entirety is decoupled from the Hubble flow.  This assumption may be problematic, however, if the Hubble flow penetrates to a radius of 1.4 Mpc or less, which is the location of our outermost tracer as shown in Figure \ref{fig:obj3}.  Assuming the potential of the LG to be Keplerian, \citet{kara2009} and \citet{courteau99} estimate the Hubble flow to begin at a spherical surface of radius 1 Mpc and 1.2 Mpc, respectively.  A more realistic treatment of the LG potential would likely yield an oddly shaped zero-velocity surface rather than a perfect sphere, and this may allow more leeway for the inclusion of bound members.  Regardless, modelling the Hubble flow at the fringes of the LG is an important consideration for improving the present work.


\section{Total Mass of the Local Group} \label{sec:lgmass}

\begin{figure}
   \begin{center}
   \includegraphics[width=0.45\textwidth]{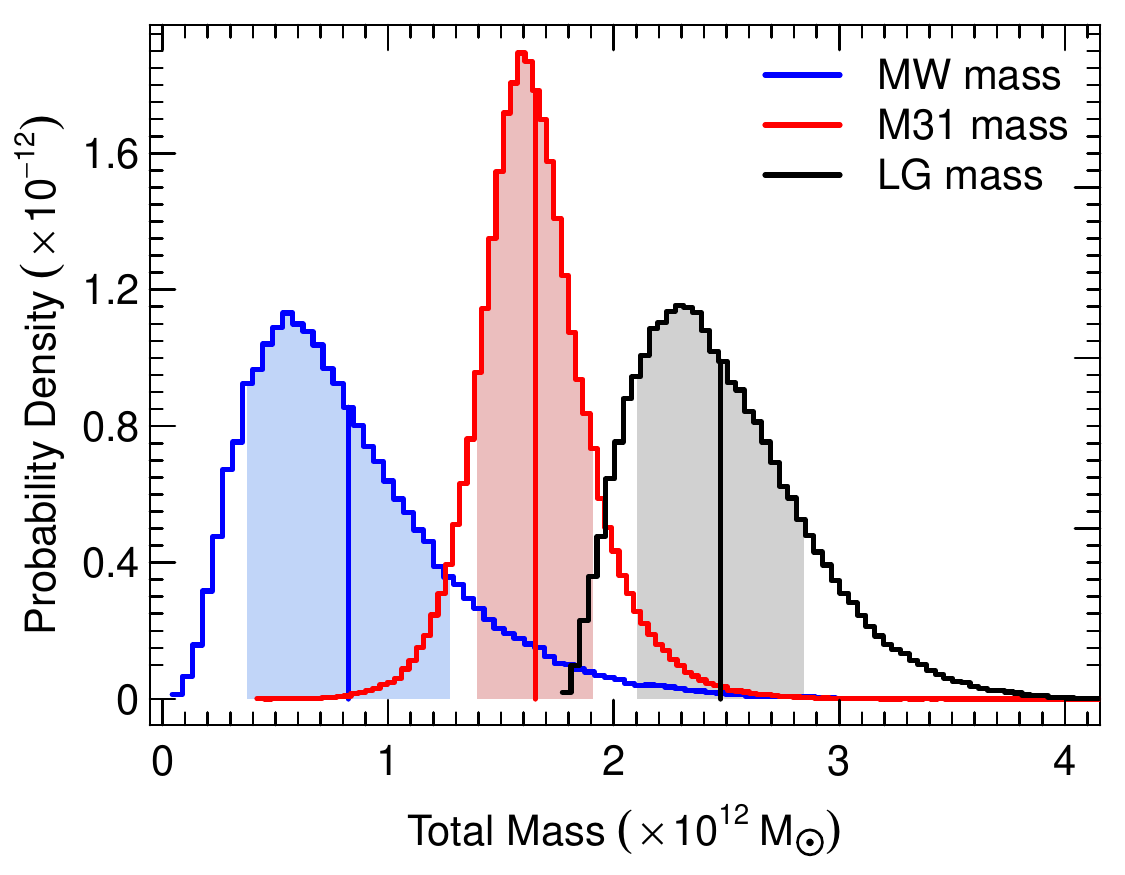}
   \caption{Probability distributions for the total mass of the LG (black), M31 (red), and the MW (blue) as calculated in Section \ref{sec:lgmass}.  The mean of each distribution is shown as a solid line, and the 1$\sigma$ uncertainty is shaded.  The skewness in these distributions comes from two sources as discussed in the text: the velocity dispersion of the outer LG galaxies, which appears as $\sigma$ in equation (\ref{eq:mest0}), and the mass ratio of the MW and M31, which appears as $m$ in equations (\ref{eq:mest0}) and (\ref{eq:mest2}).}
   \label{fig:mfinal}
   \end{center}
\end{figure}

\subsection{Method using the Virial Theorem}

A group or cluster of galaxies is virialized when the following condition holds
\begin{equation} \label{eq:vircon}
   0 = \frac{ {\rm d^2} }{ {\rm d} t^2} \sum_{i=1}^{N} M_i r_i^2,
\end{equation}
where $M_i$ is the mass of each galaxy and $r_i$ is the distance from the center of the group (\citealt{BT}, chapter 4.8.3).  Given that the MW and M31 are the dominant masses in the LG, the applicability of (\ref{eq:vircon}) reduces to asking whether the two-body interaction of the MW and M31 has settled into an equilibrium configuration.  It clearly has not, which means that the LG is not virialized\footnote{The virial theorem may apply to M31 and the MW if we average their kinetic and potential energies over an entire orbital period.  However, this requires an assumption for the orbit, in which case the timing argument may be preferrable to the virial theorem (\citealt{dlb81,kochanek96}).}.

Nevertheless, the virial theorem may plausibly apply if we consider the outer LG galaxies only.  Taking $N$ remote galaxies of similar (average) mass $M_0$, equation (\ref{eq:vircon}) says that their average squared distance must have a vanishing second derivative.  That is, the set of outer galaxies is allowed to be in a state of expansion or contraction with respect to the LG centre-of-mass, but this motion must be constant, i.e. non-accelerating.  Without any evidence to the contrary, we assume that this condition for virialization is true.  Our strategy is to estimate the LG mass from the velocity dispersion of the outer members, which requires us to represent the gravitational influence of the MW and M31 as external potentials.

To begin, we take the trace of the tensor virial theorem as given in \citet{BT} (chapter 4, equations 247, 311) to yield
\begin{equation}
   2K + W + V = 0,
\end{equation}
where $K$ is the total kinetic energy of the outer LG galaxies, $W$ is their self-gravitating potential energy, and $V$ is the potential energy arising from external gravitational fields.  Because we take as our virialized system the outer LG galaxies only, the kinetic energy is
\begin{equation} \label{eq:ke}
   K = \frac{3}{2}M_0 N \sigma^2,
\end{equation}
where $\sigma$ is the one-dimensional velocity dispersion of the system, and the factor of 3 accounts for the total dispersion under the assumption of isotropy.  We take $W\approx0$ because the self-gravity of the outer LG galaxies is negligible in comparison to the external influence of the MW and M31.  For $V$ we use
\begin{equation}
V = - M_0 \sum_{i=1}^{N} \bmath{r}_i \cdot \nabla \Phi_{\rm ext},
\end{equation}
where the external potential $\Phi_{\rm ext}$ is to be evaluated at the position $\bmath{r}_i$ of each galaxy in the system.  In parametrizing $\Phi_{\rm ext}$ we avoid NFW halos because their total mass does not converge (\citealt{nfw97}).  Instead, we use Hernquist potentials and follow the common practice of choosing the parameters to mimic comparable NFW halos (e.g. \citealt{volker2005}).  The external potential is decomposed into contributions from the MW and M31 as
\begin{equation}
\Phi_{\rm ext} (\bmath{r}) = \frac{ -{\rm G} M_{\rm MW} }{|\bmath{r} - \bmath{r}_{\rm MW}| + a_{\rm MW} } - \frac{{\rm G} M_{\rm M31} }{ |\bmath{r} - \bmath{r}_{\rm M31}| + a_{\rm M31} },
\end{equation}
where we have represented the halo potential of the MW (M31) with a Hernquist sphere of total mass $M_{\rm MW}$ ($M_{\rm M31}$), scale length $a_{\rm MW}$ ($a_{\rm M31}$), and centered at the position vector $\bmath{r}_{\rm MW}$ ($\bmath{r}_{\rm M31}$). 

Denoting the mass ratio as $m=\mrat$ and assuming that the total mass of the LG is $M_{\rm LG} = M_{\rm MW} + M_{\rm M31}$, the above equations can be combined to yield the mass estimator
\begin{equation} \label{eq:mest0}
M_{\rm LG} = \frac{- 3N (1+m) \sigma^2 }{ {\rm G} \sum_{i=1}^{N} \bmath{r}_i \cdot \nabla_i \left( p_i + m q_i \right) },
\end{equation}
where the subscript on $\nabla_i$ indicates that derivatives are to be taken with respect to coordinates $\bmath{r}_i$, and where
\begin{eqnarray} \label{eq:mest1}
p_i &=& (|\bmath{r}_i - \bmath{r}_{\rm MW}| + a_{\rm MW})^{-1}, \nonumber \\
q_i &=& (|\bmath{r}_i - \bmath{r}_{\rm M31}| + a_{\rm M31})^{-1}.
\end{eqnarray}

\subsection{Probability distributions for $\bmath{M_{\rm LG}}$, $\bmath{M_{\rm MW}}$, and $\bmath{M_{\rm M31}}$}

We must remove the MW and M31 from our preferred set of galaxies from the analysis of Section \ref{sec:res}, leaving us with $N=15$ outer galaxies.  Given this set, we compute the velocity dispersion $\sigma$ as the standard deviation of their velocities $V_{\rm LG}$ corrected to the LG rest frame.  To calculate the velocities $V_{\rm LG}$, we follow the prescription given by equation (\ref{eq:vlosc}), but we use the full MCMC chains for $\vsunlg$ (Set B) rather than the mean values.  The resulting value $\sigma= 59.7 \pm 4.1 ~\kms$ differs in two important ways from the values listed in Table \ref{tab:res1}: the motions of M31 and the MW are excluded, and the final solution for the LG rest frame is utilized rather than the intermediate solution of Step 1.

To evaluate equation (\ref{eq:mest0}), we choose the origin of our coordinate system to lie at the LG centre-of-mass, given by equation (\ref{eq:rlg}) with the adopted value $m=2.30$.  This is the natural choice of origin for which equation (\ref{eq:vircon}) may hold.  The cosmologically preferred range of virial radii for the MW and M31 is 200 to 300 kpc, and a reasonable range of halo concentrations is 10 to 20 (\citealt{klypin2002,alis2012cvir}).  Following equation 11 in the Appendix of \citet{vdm2012}, the corresponding range of values for the Hernquist scale length $a$ is roughly 20 to 60 kpc.  We split this range and choose $a_{\rm MW} = a_{\rm M31} = 40~{\rm kpc}$ in equation (\ref{eq:mest1}).

We calculate $M_{\rm LG} = (2.5 \pm 0.4) \times 10^{12}~\Msun$, quoted as the mean value and the standard deviation.  The full probability distribution for $M_{\rm LG}$ is noticeably skew and is shown as the black curve in Figure \ref{fig:mfinal}.  To determine this distribution, our final MCMC chains from Set B for the mass ratio $m$ and the solar motion $\vsunlg$ have been used in equation (\ref{eq:mest0}).  The skewed shape of the $M_{\rm LG}$ distribution is due to the velocity dispersion $\sigma$, which follows a lognormal distribution.

To estimate the total masses of the MW and M31, we assume that the entire LG mass can be divided between the MW and M31 halos.  We will see in the following discussion that this assumption is well motivated.  Taking once again the full MCMC chains for the mass ratio $m=\mrat$, we use the relations
\begin{equation} \label{eq:mest2}
M_{\rm MW} = \frac{1}{1+m} M_{\rm LG},
\qquad
M_{\rm M31} = \frac{m}{1+m} M_{\rm LG},
\end{equation}
to measure $M_{\rm MW}=(0.8 \pm 0.5) \times 10^{12}~\Msun$ and $M_{\rm M31}=(1.7 \pm 0.3) \times 10^{12}~\Msun$, where the mean and standard deviation have been quoted.

The full probability distributions are given in Figure \ref{fig:mfinal}, which shows that $M_{\rm MW}$ is strongly skewed whereas $M_{\rm M31}$ is not.  This is explained as follows.  The lognormal distribution of the mass ratio $m$ (e.g. Figure \ref{fig:mrat}) causes the quantity $1/(1+m)$ to be skewed toward small values (i.e. toward 0) whereas the quantity $m/(1+m)$ is skewed toward larger values (i.e. toward 1).  Because the distributions of $1/(1+m)$ and $M_{\rm LG}$ are both skewed toward smaller values, their product in equation (\ref{eq:mest2}) causes $M_{\rm MW}$ to have an augmented skewness.  In contrast, the quantities $m/(1+m)$ and $M_{\rm LG}$ are skewed in opposing directions, such that their product in equation (\ref{eq:mest2}) leaves $M_{\rm M31}$ with a roughly symmetric shape.

\subsection{Remarks on the mass estimates}

The virial theorem was used previously by \citet{courteau99} (hereafter \citetalias{courteau99}) to estimate the mass of the LG, but they made several simplifying assumptions. In contrast to our analysis, \citetalias{courteau99} consider all LG members in their calculations including the MW and M31. It is not immediately clear that the steady-state virial theorem holds for this system. Also, as M31 and the MW are so much more massive than the LG dwarfs, the kinetic energy should be evaluated as a mass-weighted sum rather than via equation (\ref{eq:ke}). The potential energy used in \citetalias{courteau99} may also be problematic, as it relies on the poorly known half-mass radius of the LG.  Nonetheless, \citetalias{courteau99}'s mass estimate for $M_{\rm LG}$ is $(2.3 \pm 0.6) \times 10^{12}~\Msun$, which is comparable to ours, even though rather different assumptions underlie their calculations.

The uncertainty in our measurement for $M_{\rm LG}$ is remarkably small, being up to four times smaller than the uncertainty in other recent $M_{\rm LG}$ estimates in the literature (\citealt{partridge2013,vdm2012}).  The random error arising from equation (\ref{eq:mest0}) is largely determined by the uncertainty in $\sigma$, which in turn depends on the precision of our final estimate of $\vsunlg$ (Table \ref{tab:res1}).  The uncertainty in the mass ratio $m$ is also present in our final inference, but it has a more pronounced effect for $M_{\rm MW}$ and $M_{\rm M31}$ as compared to $M_{\rm LG}$.

Because the outer LG galaxies lie at large distances from the MW and M31 ($>$350 kpc), changing the scale lengths $a_{\rm MW}$ and $a_{\rm M31}$ in equation (\ref{eq:mest1}) has only a minor effect.  Taking $a_{\rm MW}=a_{\rm M31}=20$ kpc rather than 40 kpc reduces the value of $M_{\rm LG}$ by only $1.5 \times 10^{11}~\Msun$, and taking 60 kpc increases the value by the same amount.  These effects are small in comparison to the measured uncertainties.

Our results vary somewhat when we include or exclude various members in our set of remote LG galaxies.  For instance, if we exclude Andromeda 16 from our analysis, the value of $M_{\rm LG}$ decreases by roughly $5 \times 10^{11} ~\Msun$ and the mean value of $\mrat$ drops from 2.30 to 1.78.  Figure \ref{fig:obj3} indicates the peculiarity of Andromeda 16: it is located close to the LG centre-of-mass (roughly 100 kpc away) and has a relatively large negative velocity, such that its exclusion causes the velocity dispersion of our LG galaxies to decrease by $\approx5 \kms$.  The corresponding decrease in kinetic energy per equation (\ref{eq:ke}) underlies the decrease in the value of $M_{\rm LG}$.  Andromeda 16 is therefore to the LG what Leo I is to the MW (e.g. \citealt{mbk2013}): a galaxy whose unique kinematics exerts a strong influence on the overall mass budget.

Our measurement of $M_{\rm LG}$ likely underestimates the full uncertainty because we do not incorporate any systematic errors related to our assumptions of virialization and isotropic velocities.  At the outer edge of the LG, orbits may be radially anisotropic due to the infall of objects onto the group, but this net inward motion may be compensated by outward expansion in the Hubble flow.  If we were to correct for radial anisotropy, our estimate of $M_{\rm LG}$ would decrease because the one-dimensional velocity dispersion $\sigma$ largely probes the radial direction in the LG.  Consequently, because we ignore issues of anisotropy, our adopted estimate of $M_{\rm LG}$ may be an upper limit.

For LG members located at much smaller distances, other complications may arise from the gravitational influence of the MW and M31.  Nevertheless, our method may be insensitive to some of these issues.  For instance, even though the satellites are clearly not virialized, let us apply equation (\ref{eq:mest0}) to the set of all $N=72$ LG dwarf galaxies (satellites plus outer LG).  This yields a remarkably similar answer to our adopted result: $M_{\rm LG} = (2.7 \pm 0.1 ) \times 10^{12}~\Msun$.  This is because the gain in kinetic energy ($\sigma=103.6 \pm 1.8 ~\kms$) is balanced by an increase in potential energy, since the satellites penetrate deeper into the halos of M31 and the MW.  

Our value for $M_{\rm LG}$ is inconsistent with recent estimates that use the timing argument, which give much larger values of $M_{\rm LG}=(4.93 \pm 1.63) \times 10^{12}~\Msun$ (\citealt{vdm2012}) and $M_{\rm LG}=(4.73 \pm 1.03) \times 10^{12}~\Msun$ (\citealt{partridge2013}).  This discrepancy can be addressed by the findings of \citet{gonzalez2013}, who apply the timing argument to LG analogues in $\Lambda$CDM simulations.  They conclude that when pairs are selected to match the relative velocity of the MW and M31, the total mass is overestimated by a factor of $\sim$1.6.  Applying this correction to the measurements of \citet{vdm2012} and \citet{partridge2013} yields $M_{\rm LG} \approx 3 \times 10^{12}~\Msun$, which is roughly consistent with our results.

Because the literature values for $M_{\rm MW}$ and $M_{\rm M31}$ each scatter in the range $0.5-2 \times 10^{12} ~\Msun$ (see references in Section \ref{sec:int}), our measurement of $M_{\rm LG}$ is broadly consistent with the sum of the total masses of M31 and the MW.  It is notable that our estimate of $M_{\rm MW}=(0.8 \pm 0.5) \times 10^{12}~\Msun$ is on the low end of the range of literature values.  Comparing to the mass within 50 kpc of the MW ($4 \times 10^{11}~\Msun$, \citealt{alis2012cvir}) and within 80 kpc ($7 \times 10^{11}~\Msun$, \citealt{gnedin2010}), our measurement implies that the MW halo is rather concentrated and contains little mass in its outskirts.  A similar conclusion is reached by \citet{alis2012mass} by analyzing distant BHB stars in the halo.  An even smaller mass for the MW has been put forward by \citet{gibbons2014}, who infer a total mass of only $\sim$$0.5 \times 10^{12}~\Msun$ using dynamical models of the Sagittarius Stream.

Our value for $M_{\rm LG}$ is sufficiently small that it is difficult to accommodate much mass in the LG outside of M31 and the MW.  This comes in contrast to the expectations of $\Lambda$CDM, where the build-up of the MW and M31 halos by accretion causes the entire LG to be littered with mass.  For instance, \citet{gonzalez2013} analyse a statistically significant sample of $\Lambda$CDM analogues of the LG, and they estimate an average total mass of $4.2^{+3.4}_{-2.0} \times 10^{12} ~\Msun$ within 1 Mpc, whereas the sum of the virial masses of the two halos is only $2.40^{+1.95}_{-1.05} \times 10^{12} ~\Msun$.

Because the outer LG galaxies are located at large distances from the MW and M31, their motions are sensitive dynamical probes of the mass content of the LG as a whole. However, perturbations from external mass concentrations may also be important.  It is unclear to what degree the internal dynamics of the LG is affected by external agents, but investigations have been carried out regarding orbit integrations (\citealt{peebles2011}) and mass estimates (\citealt{phelps2013}).  These studies considered the influence of nearby galaxies whose mass is on par with that of the MW and M31 (e.g. Maffei group, Sculptor group, M81, Centaurus A), but it is also relevant to consider the influence of the Virgo cluster, which although farther away is several orders of magnitude more massive.  Because we have treated the LG as an isolated system in the present work, improving upon our simple assumptions will be a nontrivial task for the future.

Nevertheless, if the outskirts of the LG are a significant reservoir of mass as suggested by $\Lambda$CDM simulations, the dispersion of the outer LG members should betray its presence.  Given that we have derived a small mass for the LG, it would seem that the outskirts of the LG are rather empty, in contrast with cosmological expectations.

\section{Discussion} \label{sec:dis}

\subsection{Relaxing the assumptions} \label{sec:relax}

\begin{figure}
   \begin{center}
   \includegraphics[width=0.4\textwidth]{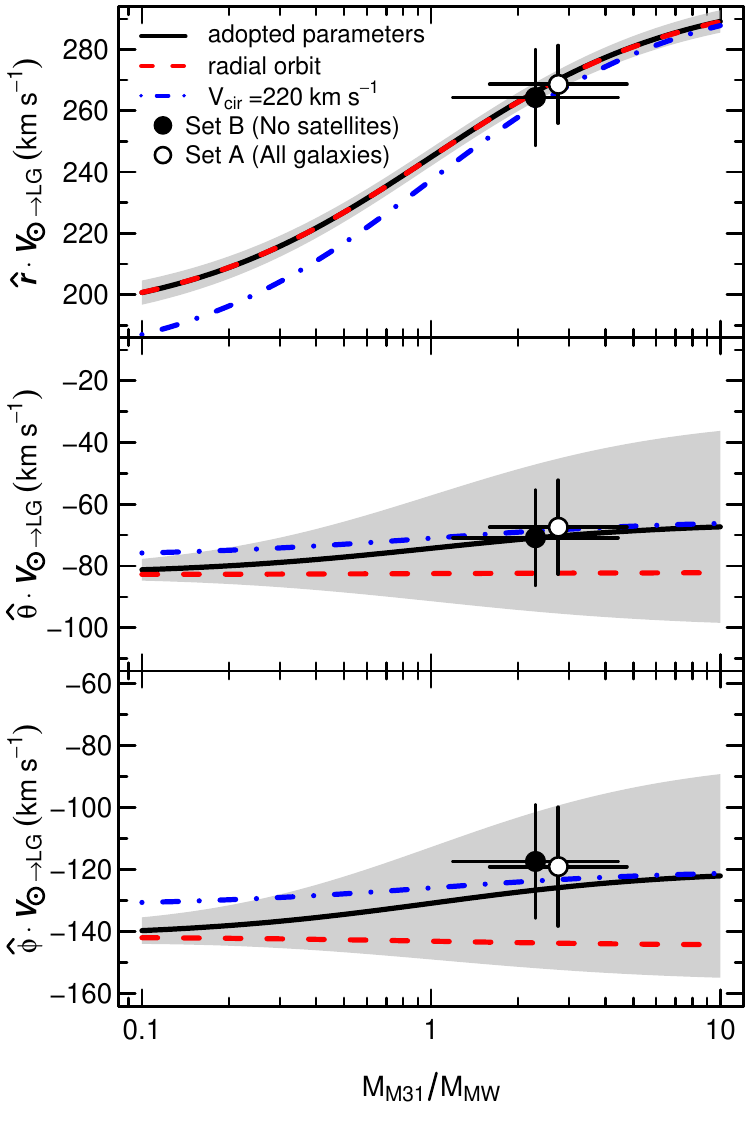}
   \caption{The solar motion $\vsunlg$ as projected along $\rhat$ (top), $\that$ (middle), and $\phat$ (bottom) as a function of the mass ratio $\mrat$.  As described in Section \ref{sec:relax}, the solid black line is determined by momentum balance and the chosen parameters of Section \ref{sec:tbg2}.  The grey shaded region gives the $1\sigma$ uncertainty, which is much larger in the tangential components due to the large uncertainty in the M31 transverse (heliocentric) motion.  The other lines give the solution for different choices of parameters: M31 tangential velocity corresponding to a radial orbit (red dashed line), and MW circular velocity of $\vcir=200~\kms$ (blue dot-dashed line).  We can regard our analyses of Sets A and B as a way to select the preferred region along these curves.  The corresponding final measurements for $\mrat$ and $\vsunlg$ are given by circles (A white, B black), and their $1\sigma$ uncertainties are given as bars.}
   \label{fig:last-vsunrpt}
   \end{center}
\end{figure}

Our inference of $\mrat$ is fundamentally limited by the uncertainty in $\vsunlg$.  It is therefore worthwhile to turn the problem around and ask whether $\vsunlg$ can be determined by an assumed mass ratio.  Given our choices for the heliocentric velocities of M31 and the MW in Section \ref{sec:tbg2}, we can use equations (\ref{eq:mom}) and (\ref{eq:decomp}) to define the vector $\vsunlg$ as a function of the mass ratio.  Figure \ref{fig:last-vsunrpt} shows the resulting values for $\vsunlg$ as projected along $\rhat$, $\that$, and $\phat$ for $\mrat$ ranging from 0.1 to 10.

Figure \ref{fig:last-vsunrpt} shows clearly that the radial component of $\vsunlg$ is very sensitive to the value of the mass ratio, as it changes by almost $100~\kms$ over the range $\mrat=0.1$ to 10.  On the other hand, the mean values of the tangential components are comparatively insensitive, changing by only $\sim$$20~\kms$ over the same range.  In other words, the radial component of $\vsunlg$ essentially determines the mass ratio.

These trends do not change appreciably when we pick different parameter values for the MW circular velocity and the M31 transverse motion.  The dash-dotted blue curve in Figure \ref{fig:last-vsunrpt} gives the mean result when we decrease the value of $\vcir$ from $239~\kms$ to $220~\kms$.  It is only slightly different from our adopted case (black curve), departing at most by $\sim$$10~\kms$.  A different choice for the M31 transverse motion is given by the red dashed curve, which corresponds to the purely radial orbit.  Once again, the difference is minor.  It is interesting that these different parameter values actually accentuate the previously noted dependence on the mass ratio: the radial component of $\vsunlg$ steepens as a function of the mass ratio, and the tangential components flatten.

If we think of the black curves in Figure \ref{fig:last-vsunrpt} as theoretical constraints, we can regard the main analysis of this paper as a way to select the preferred location along the curves.  The results of our analysis for Sets A and B are shown as white and black circles, respectively, in Figure \ref{fig:last-vsunrpt}, and as required they follow the trends given by the curves.

Our analysis as outlined in Sections \ref{sec:meh1} and \ref{sec:meh2} may be repeated with other combinations of LG members motivated by different selection criteria.  However, it is not clear that an ideal set of independent LG members exists.  Though we excluded satellites from our analysis, one may wish to go even further and exclude M31 and the MW themselves from Step 1 of our method.  This would leave only the outer LG tracer population to determine the bulk motion of the LG, prior to updating the posteriors in Step 2 by balancing the momentum of M31 and the MW.

Such an approach may be attractive, since the MW and M31 are not bona fide tracers owing to their considerable masses.  We performed this alternate analysis but found very little difference with our main results for Set B, as summarized in Tables \ref{tab:res1} and \ref{tab:res2}.  For instance, this analysis yields a mass ratio given by $\log_{10} (\mrat) = 0.40 \pm 0.30$ and a solar motion given by $\vsun = 299.0 \pm 14.9 ~\kms$, $\lsun = 99.1^{\circ} \pm 4.0^{\circ}$, and $\bsun = -6.3^{\circ} \pm 3.4^{\circ}$.  We therefore find no reason to stray from our fiducial analysis of Set B.

\subsection{Arguments for $\bmath{\mrat>1}$} \label{sec:mrat}
In this section we use a number of independent arguments to interpret M31 as the most massive member of the LG.  We compare M31 and the MW in terms of their maximum rotational velocities, their satellite and globular cluster populations, their luminosities, and their stellar content.

\subsubsection{Maximum rotational velocities}

Considering that the rotation curves of spiral galaxies are remarkably flat, the (luminous) matter which dominates the inner region must be related in a fundamental way to the (dark) matter which dominates the outer region.  This fact can be used to estimate dynamical masses of spiral galaxies from their rotation curves, by way of the well-known Tully-Fisher relation (\citealt{tully77}): $\mrat \simeq \xi^4$, where $\xi$ stands for the ratio of maximum rotational velocities in the disks of M31 and the MW.  The lack of strict equality intends to show that this relationship is not precise and has a certain amount of intrinsic scatter (\citealt{edlb82}).

Taking the maximum rotational speed of the MW in the range 220-250 $\kms$ (\citealt{kdlb86,reid&brunthaler,reid2009}) and taking the range 250-260 $\kms$ for M31 (Figures 7, 8 of \citealt{corbelli2010}), we estimate a minimum value for $\mrat$ of 1.0 (taking $\xi = 250/250$), a maximum value of 2.0 ($\xi = 260/220$), and a likely value of 1.2 ($\xi = 250/240$).

\subsubsection{Dwarf satellites and globular clusters}

\begin{figure*}
   \begin{center}
   \includegraphics[width=1.0\textwidth]{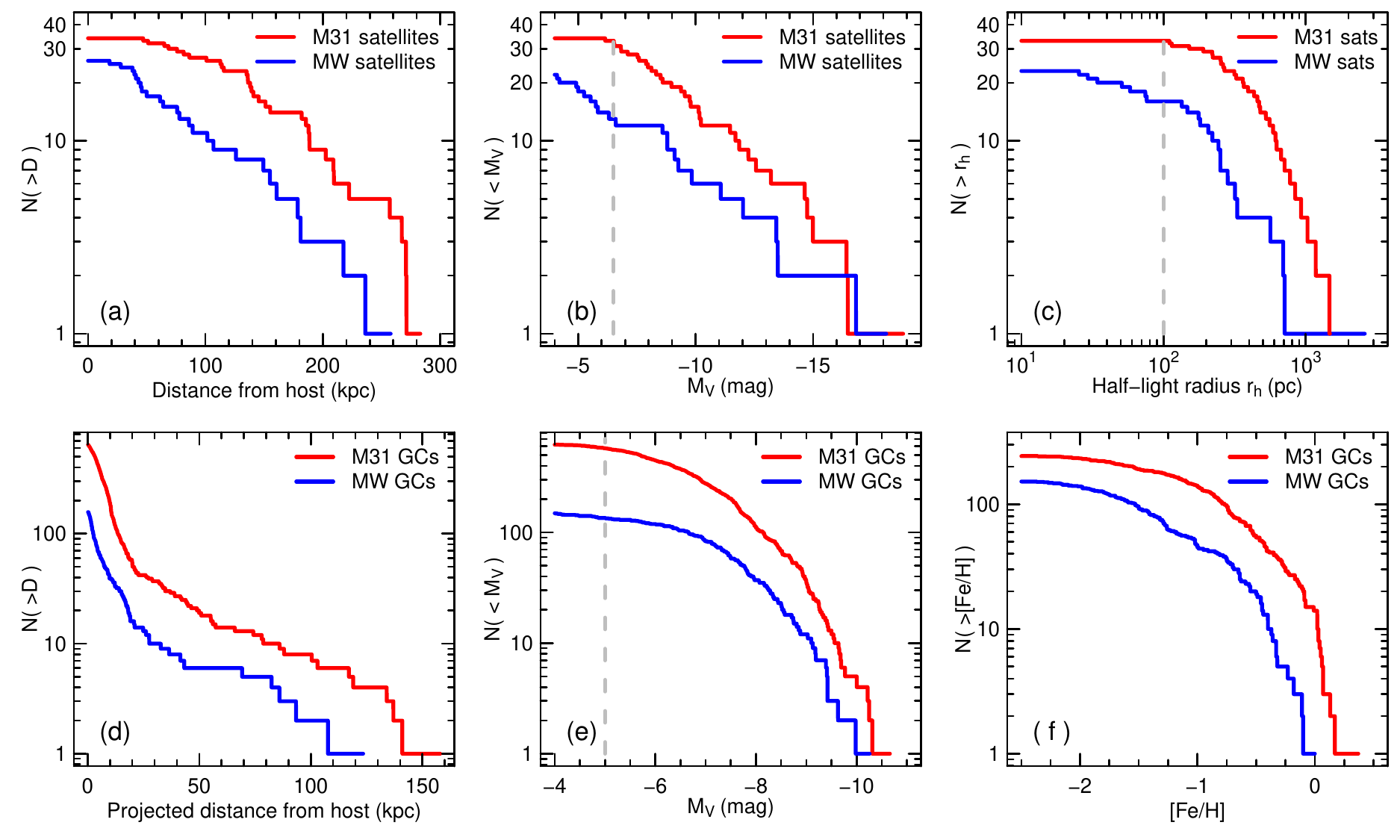}
   \caption{A comparison of the dwarf satellites and globular clusters (GCs) of M31 (red lines) and the MW (blue lines), where each panel gives the cumulative number of objects that pass above a running threshold on the $x$-axis.  The top row shows data for the dwarf satellites, taken from \citet{mcconn2012} with the selection cut of being within 350 kpc of the MW or M31.  We exclude Andromeda 16 because its distance from M31 can exceed 350 kpc to within 1$\sigma$ error, and we exclude Canis Major as it may be a feature of the MW disk rather than a dwarf galaxy.  Panel (a) gives the distance from the host, panel (b) gives the absolute visual magnitude $M_V$, and panel (c) gives the half-light radius $r_h$.  The bottom row shows data for the GCs, taken from the updated catalogs of \citet{harris96} (December 2010 edition) for the MW and \citet{galleti2004} (version 5, August 2012) for M31.  In addition, we include data on 20 recently discovered GCs in the outer halo of M31 (\citealt{mackey2013,zinn2013}).  Because the distances to individual GCs of M31 are not measured, panel (d) gives the projected distance from the host, rather than the true distance.  To maintain a consistent comparison, we project the locations of the MW GCs onto the $\that$-$\phat$ plane (i.e. the same plane for viewing the M31 GC system).  Panel (e) gives the absolute visual magnitude $M_V$, where the apparent magnitudes of the M31 GCs are transformed to absolute magnitudes using the distance modulus of M31.  Panel (f) gives the metallicity of the GCs, where only $\sim$1/3 of the M31 GCs have measured (spectroscopic) metallicities (\citealt{galleti2009}).  The grey dashed lines in panels (b), (c), and (e) give the rough locations for the completeness threshold.}
   \label{fig:sats-gcs}
   \end{center}
\end{figure*}

Figure \ref{fig:sats-gcs} compares the population of satellite galaxies and globular clusters (GCs) surrounding M31 and the MW.  The panels give the cumulative number of objects that pass above a running threshold on the $x$ axis for various quantities, including absolute magnitude, half-light radius, metallicity, and distance (or projected distance) from the host galaxy.  The main conclusion to be drawn from Figure \ref{fig:sats-gcs} is that the M31 population of satellites and GCs is more extended and has more total members than the MW population.  In addition, the M31 population generally has an excess of large, bright, and metal-rich members.  As discussed below, these are indications that the halos of M31 and the MW differ in a fundamental way, and we infer that M31 is both more massive and more extended. 

As shown in panel (a) of Figure \ref{fig:sats-gcs}, the M31 dwarf galaxies are located at larger distances from their host as compared to the MW satellites.  Whether or not this discrepancy can be addressed by incompleteness issues is unclear (\citealt{yniguez2014}), but taken at face value it would likely imply that the M31 halo is larger.  As shown in panel (b), the satellites of M31 outnumber those of the MW at most visual magnitudes, although there is an overlap of the curves at the bright end.  This is due to the Magellanic Clouds, which are actively forming stars and are abnormally blue for their mass (\citealt{tollerud2011}).

The overlap in panel (b) can be mitigated by instead considering $K$-band luminosities (e.g. \citealt{kara2013}), which is a better diagnostic for the mass content of galaxies.  Accordingly, it would seem that the M31 satellite system is a larger reservior of stellar and possibly also total mass.  The physical sizes of the M31 satellites are also generally larger than their MW counterparts, as given by their half-light radii in panel (c).  The MW curve in panel (c) extends beyond the M31 curve for one galaxy only, the Sagittarius dwarf.  It has the largest half-light radius of any satellite because it is in the process of being pulled apart by the tidal field of the MW.

In total there are 157 GCs in the MW dataset and roughly four times as many (644) in the M31 dataset.  This fact alone supports M31 as being more massive, because the specific frequency of GCs is indicative of the total mass within a galaxy (\citealt{peng2008,harris2013}).  In fact, \citet{hudson2014} find that GC populations form in direct proportion to the total halo mass of their host galaxy, which implies M31 is more than three times as massive as the MW.  The M31 GCs are also more numerous at all radii as seen in panel (d) of Figure \ref{fig:sats-gcs}, and they extend farther into the halo than the MW GCs (see also \citealt{huxor2014}).  However, it is almost certain that particularly faint or distant GCs remain undetected in both galaxies, so it is not clear whether future discoveries of outlying GCs will favor the MW or M31.

The MW GC population is outnumbered at every value of visual magnitude (panel e) and metallicity (panel f).  In particular, M31 possesses more metal-rich GCs, such that the most metal-rich GC in the MW would only be the 15th most enriched in M31.  Because GCs are devoid of dark matter, their metallicity likely correlates with the mass of their host galaxy.  Consequently, the relative excess of metal-rich GCs in M31 provides further support for $\mrat>1$.

\subsubsection{Luminosities and stellar content}

M31 is estimated to be a factor $\sim$1.3 times as luminous as the MW by \citet{vdb99}, but this is without corrections to internal absorption.  Owing to the high inclination of the M31 disk on the sky, absorption by dust in the M31 disk can significantly attenuate the total luminosity that we observe.  \citet{tempel2010} estimate that up to 20\% of the total $B$-band luminosity may be obscured from us.  Accordingly, the total luminosity of M31 could easily be a factor two to five times larger than that of the MW.

In their recent catalog of the Local Volume, \citet{kara2013} apply extinction and absorption corrections to derive absolute $B$-band magnitudes for all galaxies within 11 Mpc.  Of these galaxies, there are only two which are as bright or brighter than M31 ($M_B = -21.4$) whereas there are fifteen that are as bright or brighter than the Milky Way ($M_B = -20.8$).

A similar discrepancy in absolute magnitude holds in the $K$-band.  \citet{hammer2007} apply dust extinction corrections to the \textit{Spitzer} data of \citet{barmby2006} and determine $M_K = -24.7$ for M31, which is significantly brighter than the Milky Way at $M_K = -24.02$, as determined from \textit{COBE} data (\citealt{drimmel2001}).  Because $K$-band luminosity is a reliable proxy for total stellar mass, M31 clearly dominates the Milky Way in terms of total light and total stellar content.

The accretion histories of these two spiral galaxies also appear to be markedly different.  Noting that the MW is relatively deficient in stellar mass, angular momentum, and metal enrichment in its outskirts, \citet{hammer2007} argue that the MW likely avoided any significant merger in the last $\sim$10 Gyr.  This picture is echoed by the findings of \citet{alis2013}, who link the sharp density fall-off of the MW stellar halo to a quiescent accretion history.  The M31 stellar halo in contrast has a smooth profile (\citealt{ibata2014}), suggestive of a more active and prolonged accretion history.  We infer that M31 has been more successful than the Milky Way in assimilating dwarf galaxies over its lifetime, but whether this is due to a larger gravitational potential or simply a difference in environmental factors is unclear.

Under the assumption that the stellar mass within a galaxy can be matched monotonically to the total mass within its halo, \citet{guo2010} determine that the halo of M31 is 1.5 times more massive than that of the MW.  This does not contradict the possibility that the MW outweighs M31, however, because neither luminosity nor stellar content are exact analogs for dynamical mass.  We leave it to the reader to decide whether the camel's back is strong enough to withstand these few added straws.

\section{Summary} \label{sec:sum}

In this paper we have measured the mass ratio $\mrat$, the solar motion with respect to the LG centre-of-mass $\vsunlg$, and the total mass of the LG $M_{\rm LG}$.  The combination of our measurements for the mass ratio and $M_{\rm LG}$ allows us to estimate the individual masses of the MW and M31.  Our analysis is enabled by the well-motivated assumption that M31 and the MW have equal and opposite momenta.

Our method is a Bayesian procedure of two steps.  In Step 1, we use the collective motions of the independent (i.e. non-satellite) members of the LG to measure $\vsunlg$.  In Step 2, we impose momentum balance to estimate the velocities of the MW and M31 with respect to the LG.  The ratio of these velocities delivers the mass ratio $\mrat$.  The measurement of $\vsunlg$ is then updated in a Bayesian way to be consistent with momentum balance.  To estimate $M_{\rm LG}$, we calculate the velocity dispersion of the outer LG galaxies and apply the virial theorem (Section \ref{sec:lgmass}).

The best sample of galaxies to use in our analysis is a set of independent members of the LG. The satellites of the MW and M31 are inadmissible, as their motion is heavily influenced by their hosts. Accordingly, our preferred sample is the MW, M31, and all outlying members of the LG at least 350 kpc distant from the MW and M31.  Using this sample, we find that the quantity $\log_{10} (\mrat)$ is normally distributed with a mean and standard deviation of $0.36 \pm 0.29$, which corresponds to $\mrat = 2.3_{-1.1}^{+2.1}$.  The rest frame of the LG is given by the solar motion $\vsunlg$, which has an amplitude $V_{\odot} = 299 \pm 15 {\rm ~km ~ s^{-1}}$ in a direction towards galactic longitude $\lsun=98.4^{\circ} \pm 3.6^{\circ}$ and galactic latitude $\bsun=-5.9^{\circ}\pm 3.0^{\circ}$.

The total LG mass is calculated by applying the virial theorem to the outer members of the LG, accounting for their potential energy in the external gravity field of the MW and M31. This gives $M_{\rm LG} = (2.5 \pm 0.4) \times 10^{12}~\Msun$, where the quoted uncertainty includes random errors only and is dominated by the uncertainty in the velocity dispersion. Combined with our measurement of the mass ratio, the individual masses of the MW and M31 are $M_{\rm MW}=(0.8 \pm 0.5) \times 10^{12}~\Msun$ and $M_{\rm M31}=(1.7 \pm 0.3) \times 10^{12}~\Msun$ respectively. The quoted values are the mean and standard deviation, but the full probability distributions are noticeably skew (Figure \ref{fig:mfinal}).

Recent applications of the timing argument yield a value for $M_{\rm LG}$ which is twice as large as our estimate (\citealt{partridge2013, vdm2012}).  This discrepancy is mitigated by the results of \citet{gonzalez2013}, who find that the timing argument overestimates the total mass of LG analogues in $\Lambda$CDM simulations by a factor of $\sim$1.6.  If we correct for this bias, the timing argument can be brought into better agreement with our results by taking $M_{\rm LG} \approx 3 \times 10^{12}~\Msun$.  This value agrees well with the sum of individual mass estimates for the MW and M31, which implies that the entire mass of the LG is concentrated in the halos of the MW and M31.  It would seem therefore that the outskirts of the LG (up to 1.5 Mpc away) are surprisingly empty.

\section*{Acknowledgements}

We would like to extend our gratitude to Donald Lynden-Bell for a number of spirited discussions at various stages during the project.  We also thank the referee for a thoughtful critique of the paper, and we thank Ziv Mikulizky for valuable correspondence.  JDD is supported by a Gates Cambridge Scholarship.

\bibliographystyle{mn2e}
\bibliography{ms}

\end{document}